\documentclass[10pt,journal,compsoc]{IEEEtran}

\usepackage{url}
\usepackage{amssymb}
\usepackage{amsfonts}
\usepackage{amsmath}
\usepackage{amsmath,amssymb,verbatim}
\usepackage{graphicx}
\usepackage{float}
\usepackage{epsfig}
\usepackage{subfig}
\usepackage{url}
\usepackage{threeparttable}
\usepackage[lined,linesnumbered,ruled,vlined]{algorithm2e}
\usepackage{color}
\usepackage{graphicx,dblfloatfix}
\usepackage{changepage}%
\usepackage{layouts}

\ifCLASSOPTIONcompsoc
  \usepackage[nocompress]{cite}
\else
  \usepackage{cite}
\fi
\ifCLASSINFOpdf
\else
\fi
\newtheorem{theorem}{Theorem}

\newtheorem{definition}{Definition}

\floatname{algorithm}{Algorithm}

\begin{document}
%
\title{Reliable Virtual Machine Placement \\ and Routing in Clouds}
%
%
%
%

\author{Song Yang,
        Philipp Wieder,
        Ramin Yahyapour, 
        Stojan Trajanovski,
        Xiaoming Fu,~\IEEEmembership{Senior Member,~IEEE}
\IEEEcompsocitemizethanks{\IEEEcompsocthanksitem S. Yang and P. Wieder are with Gesellschaft f{\"u}r wissenschaftliche Datenverarbeitung mbH G{\"o}ttingen (GWDG), G{\"o}ttingen, Germany.\protect\\
E-mail: \{S.Yang, P.Wieder\}@gwdg.de
\IEEEcompsocthanksitem R. Yahyapour is with GWDG and Institute of Computer Science, University of G{\"o}ttingen, G{\"o}ttingen, Germany. 
E-mail: R.Yahyaour@gwdg.de
\IEEEcompsocthanksitem This work was done while S. Trajanovski was with the University of Amsterdam and Delft University of Technology, The Netherlands. S. T. is now with Philips Research and Delft University of Technology. \protect\\
E-mail: S.Trajanovski@tudelft.nl
\IEEEcompsocthanksitem X. Fu is with Institute of Computer Science, University of G{\"o}ttingen, G{\"o}ttingen, Germany. 
E-mail: Fu@cs.uni-goettingen.de
}
\thanks{A preliminary part of this paper appeared as conference publication \cite{RNDM16}.}}

\IEEEtitleabstractindextext{%
\begin{abstract}
In current cloud computing systems, when leveraging virtualization technology, the customer's requested data computing or storing service is accommodated by a set of communicated virtual machines (VM) in a scalable and elastic manner. These VMs are placed in one or more server nodes according to the node capacities or failure probabilities. The VM placement availability refers to the probability that at least one set of all customer's requested VMs operates during the requested lifetime. In this paper, we first study the problem of placing at most $H$ groups of $k$ requested VMs on a minimum number of nodes, such that the VM placement availability is no less than $\delta$, and that the specified communication delay and connection availability for each VM pair under the same placement group are not violated. We consider this problem with and without Shared-Risk Node Group (SRNG) failures, and prove this problem is NP-hard in both cases. We subsequently propose an exact Integer Nonlinear Program (INLP) and an efficient heuristic to solve this problem. We conduct simulations to compare the proposed algorithms with two existing heuristics in terms of performance. Finally, we study the related reliable routing problem of establishing a connection over at most $w$ link-disjoint paths from a source to a destination, such that the connection availability requirement is satisfied and each path delay is no more than a given value. We devise an exact algorithm and two heuristics to solve this NP-hard problem, and evaluate them via simulations.
\end{abstract}

}

\maketitle

\IEEEdisplaynontitleabstractindextext

%
\IEEEpeerreviewmaketitle

\IEEEraisesectionheading{\section{Introduction}\label{sec:introduction}}

\IEEEPARstart{C}{loud} computing \cite{Mell10} is a distributed computing and storing paradigm, which can provide scalable and reliable service over the Internet for on-demand data-intensive applications (e.g., on-line search or video streaming) and data-intensive computing (e.g., analyzing and processing a large volume of scientific data). The key features of cloud computing, including ``pay-as-you-go'' and ``elastic service'', attract many service providers and customers to deploy their workload from their own infrastructures or platforms to public or private clouds.

Distributed cloud systems are usually composed of distributed inter-connected data centers, which leverage virtualization technology to provide computing and storage service for each on-demand request. Once a request arrives, several virtual machines (VM) are created in one or more server nodes (which may be located in the same or different data centers) in order to accommodate the request. However, the server node failures caused by hardware malfunctions such as hard disk or memory module failures and software problems such as software bugs or configuration errors may result in the loss of the VMs hosted on it and hence the whole service cannot be guaranteed. An efficient way to overcome this concern is to create and place more VM replicas, but this approach should also take the nodes' availabilities into account. For instance, if all the VMs together with their replicas are placed at nodes with high failure probability, then a proper service cannot be guaranteed. The VM placement availability, a value between $0$ and $1$, is therefore important and refers to the probability that at least one set of all customer's requested VMs is in the operating state during the entire requested lifetime. 

Moreover, if two or more VMs are placed on different nodes, we should also ensure reliable communications between these VMs. In fact, a single unprotected path will fail if one of the links belonging to it fails. To increase the reliability of transporting data from a source to a destination, path protection (or survivability) is called for. For instance, by allocating a pair of link-disjoint paths from a source to a destination, the data is transported on the primary path. Upon a failure of the primary path, the data can be switched to the backup path. However, the path protection mechanism, which does not allow for more than $2$ link-disjoint paths, may still be not reliable enough and $w>2$ link-disjoint paths may be needed. Moreover, the link availability should also be taken into account.    
For a connection over at most $w$ link-disjoint paths between a node pair, its availability specifies the probability that at least one path is operational. Connection availability is therefore important to quantitatively measure the availability of delivering data between VMs located on different nodes in a cloud.

In this paper, we first study the Reliable VM Placement (RVMP) problem, which is to place at most $H$ groups of $k$ requested VMs on a minimum number of nodes, such that the VM placement availability is no less than $\delta$, and the specified communication delay and connection availability for each VM pair are not violated. 

Following that, we study the Availability-Based Delay-Constrained Routing (ABDCR) problem, which is to establish a connection over at most $w$ (partially) link-disjoint paths from a source to a destination such that the connection availability is at least $\eta$ and each path has a delay no more than $D$. Our key contributions are as follows:

\begin{itemize}
\item We propose a mathematical model to formulate VM placement availability with and without Shared-Risk Node Group failures, and prove that the Reliable VM Placement (RVMP) problem under both cases is NP-hard.
\item We propose an Integer Nonlinear Program (INLP) and a heuristic to solve the RVMP problem.
\item We compare the proposed algorithms with two existing heuristics in terms of performance via simulations.
\item We prove that the ABDCR problem is NP-hard, devise an exact algorithm and two heuristics to solve it, and further verify them.
\end{itemize}

The remainder of this paper is organized as follows: Section~\ref{Sec:RelWork} presents the related work. Section~\ref{Sec:VMReliability} and \ref{Sec:SRNG} formulate the VM placement availability calculation without and with SRNG failures, respectively. In Section~\ref{Sec:ProbDef}, we study the Reliable VM Placement (RVMP) problem and prove it is NP-hard. We propose an exact Integer Nonlinear Program (INLP) and a heuristic to solve the RVMP problem. The proposed algorithms are also evaluated via simulations. In Section~\ref{Sec:Routing}, we define the Availability-Based Delay-Constrained Routing (ABDCR) problem, prove the problem is NP-hard, and propose an exact algorithm and two heuristics to solve it. We also conduct simulations to verify the proposed algorithms as well. Finally, we conclude in Section~\ref{Sec:Conclusion}.

\section{Related Work} \label{Sec:RelWork}
A high-level comprehensive survey about VM placement can be found in \cite{Jennings14} \cite{Mann15}. 

\subsection{Network-Aware VM Placement}

Alicherry and Lakshman \cite{Alicherry12} first investigate how to place requested VMs on distributed data center nodes such that the maximum length (e.g., delay) of placed VM pairs is minimized. A 2-approximation algorithm is proposed to solve this problem when a triangle link length is assumed. They subsequently study how to place VMs on physical machines (racks and servers) within a data center in order to minimize the total inter-rack communication costs. Assuming that the topology of the data center is a tree, they devise an exact algorithm to solve this problem. Finally, they propose a heuristic for partitioning VMs into disjoint sets (e.g., racks) such that the total communication costs between VMs belonging to different partitions is minimized.

Biran \emph{et al.} \cite{Biran12} address the VM placement problem by minimizing the min-cut ratio in the network, which is defined as the used capacity of the cut links consumed by the communication of VMs divided by the total capacity of the cut links. They prove this problem is NP-hard and propose two efficient heuristics to solve it. Jiang \emph{et al.} \cite{Jiang12} jointly consider the VM placement and routing problem within one data center network. They propose an approximation on-line algorithm leveraging the technique of Markov approximation.

Meng \emph{et al.} \cite{Meng10} address the problem of assigning VMs to slots (CPU/memory on a host) within a data center network in order to minimize total network costs. They prove the problem is NP-hard and propose a heuristic that tries to assign VMs with large mutual rate requirement close to each other. 

\subsection{Reliable VM Placement}

Israel and Raz \cite{Israel13} study the Virtual Machine Recovery Problem (VMRP). The VMRP is to place the backup VMs for their corresponding servicing VMs on either active or inactive host, which needs to strike a balance between the (active) machine maintenance cost and VM recovery Service Level Agreement (e.g., recovery time). They show that the VMRP is NP-hard, and they propose a bicriteria approximation algorithm and an efficient heuristic to solve it.

Bin \emph{et al.} \cite{Bin11} tackle the VM placement problem by considering k-resiliency constraint to guarantee high availability goals. A VM is marked as $k$-resilient, if its current host fails and there are up to $k-1$ additional host failures, and it can still be guaranteed to relocate to a non-failed host. In this sense, a placement is said to be $k$-resilient if it satisfies the $k$-resiliency requirements of all its VMs. 
They first formulate this problem as a second order optimization statement and then transform it to a generic constraint program in polynomial time. 

Zhu \emph{et al.} \cite{Zhu14} address the Reliable Resource Allocation (RRA) problem. In this problem, each node has a capacity limit of storing VMs and each link is associated with an availability value (between $0$ and $1$). The problem is to find a star of a network to place the requested VMs, such that the node capacity limit is obeyed and the availability of the star is no less than the specified. They prove that the RRA problem is NP-hard and propose an exact algorithm as well as a heuristic to solve it. However, the defined problem in \cite{Zhu14} does not consider the node's availability and also it restricts to find a star instead of an arbitrary subgraph. 

Li and Qian \cite{Li15} assume that the VM reliability requirement is equal to the maximum fraction of VMs of the same function that can be placed in a rack. Yang \emph{et al.} \cite{Yang15B} develop a variance-based metric to measure the risk of violating the VM placement availability requirement, but none of them take VM replicas/backups into account. Nevertheless, none of above papers quantitatively model the availability of VM placement (and solve the respective reliable VM placement problem), as we do in this paper.  

\subsection{Availability-Aware Routing}
Song \emph{et al.}~\cite{Song07} propose an availability-guaranteed routing algorithm, where different protection types are allowed. They define a new cost function for computing a backup path when the unprotected path fails to satisfy the availability requirement. She
\emph{et al.}~\cite{She10} prove the problem of finding two link-disjoint paths with maximal reliability (availability) is NP-hard. They also propose two heuristics for that problem. Luo \emph{et al}.~\cite{Luo09} address the problem of finding one unprotected path or a pair of link-disjoint paths, such that the cost of the entire path(s) is minimized and the reliability requirement is satisfied. To solve it, they propose an exact ILP as well as two approximation algorithms. However, the reliability (availability) calculation in~\cite{Luo09} is different from the
aforementioned papers, and assumes a single-link failure model. Assuming each link in the network has a failure probability (=1-availability), Lee \emph{et al.} \cite{Lee10} minimize the total failure probability of unprotected, partially link-disjoint and fully link-disjoint paths by establishing INLPs. They further transform the proposed INLPs to ILPs by using linear approximations. Yang \emph{et al.} \cite{RNDM, Yang15} study the availability-based path selection problem, which is to find at most $w$ (partially) link-disjoint paths and for which the total availability is no less than the specified. They prove that this problem is NP-hard and cannot be approximated to an arbitrary degree when $w \geq 2$. They propose an exact INLP and a heuristic to solve this problem. 

\section{VM Placement availability} \label{Sec:VMReliability}

The availability of a system is the fraction of time that the system is operational
during the entire service time. The availability $A_{j}$ of a network component $j$ can be calculated as \cite{Mccool03}:
\begin{equation}
A_{j}=\frac{MTTF}{MTTF+MTTR}%
\end{equation}
where $MTTF$ represents Mean Time To Failure and $MTTR$ denotes Mean Time To
Repair. In this paper, a node in the network represents a server, and its availability is equal to the product of the availabilities of all its components (e.g., hard disk, memory, etc.). 
In reality, we can obtain the server's availability value by accessing the detailed logs extracting every hardware component repair/failure incident during the lifetime of the server. 
The details for characterizing server and other data center network device (e.g., switches) failures can be found in \cite{Vishwanath10} and \cite{Gill11}.
Since our focus in this paper is not on how to calculate the device's availability, we assume that the server availabilities (or the SRNG event failure probabilities) value are known. 
Moreover, we assume a general multiple node (link) failure scenario, which means at one particular time point, multiple nodes (links) may fail. In this section, we first assume that the node availabilities are uncorrelated/independent.

We assume that the user request consists of $k$ VMs with associated communication requirements (we consider delay and connection availability in this paper) between different VM pairs. These $k$ VMs are represented by $v_1$, $v_2$,\ldots, $v_k$. 
For each requested VM $v_i$ ($1 \leq i \leq k$), placing it on the same node (say $n$) more than once cannot increase placement availability, since when $n$ fails, all its resident VMs will fail simultaneously. Therefore, we need to place $v_i$ on different nodes to increase the placement availability. Let us use $H_i$ to represent the maximum number of nodes to host VM $v_i$. Or, equivalently, $H_i$ indicates the maximum number of nodes that $v_i$ can be placed on.
We denote $H=\max_{i=1}^{k}(H_i)$. We distinguish and analyze the VM placement availability under two different cases, namely (1) Single Placement: each VM is placed on exactly $H=1$ node in the network, and (2) Protected Placement: $\exists v_j \in V$, such that $v_j$ can be placed on $H_j>1$ nodes in the network, i.e., $H>1$. 
In the following, we will address the VM placement availability under two node failure scenarios, namely, (1) single node failure scenario: at most one node may encounter failure at any particular time point, and (2) multiple nodes failure: multiple nodes may fail at any particular time point. Without loss of generality, in this paper, we assume multiple node failure scenario. Moreover, we assume that the servers are heterogeneous and they can be located in either the same data center or different data centers. 

\subsection{Single-node failure}

Here it is assumed that all the nodes in the network have very low failure probability (highly reliable). Therefore, we can assume that at one time point, at most one node may encounter failure. In the single placement case, if $m$ nodes with availability $A_1$, $A_2$,\ldots, $A_m$ are used for hosting $k$ VMs ($m \leq k$), then the availability of the VM placement is $A_{sn}=\min (A_1, A_2, \ldots, A_m)$. In the protected placement case, if there are another $m'$ ($1 \leq m' \leq k$) nodes which are totally different from the existing $m$ nodes and $k$ VMs are also placed on these $m'$ nodes. In this sense, the availability of placing in total $2k$ VMs on $m+m'$ nodes is $1$, since each VM located on one node is fully ``protected'' by another backup VM located on a different node. We can also see that for each VM, one backup VM placed on a different node is enough, i.e., there is no need to have more than one backup VM. Moreover, when there are less than $k$ backup VMs placed on $m'$ nodes, it indicates that at least one VM does not have its backup. Let us denote the node set $\mathcal{N}_{un}$ as the nodes on which VMs are located and do not have their backups. As a result, the availability of placing $g$ ($k<g<2k$) VMs on $m+m'$ nodes is $\min_{i \in \mathcal{N}_{un}} (A_i)$. However, this approach only works when all the links are highly reliable. In Appendix A, we will provide an Integer Nonlinear Program (INLP) to solve the Reliable Virtual Machine Placement problem under the single-node failure scenario.

\subsection{Multiple node failure}
It is a more general model where all the nodes may fail simultaneously at any particular time point. In this context, the VM placement availability in the single placement is equal to the product of the availabilities of nodes that host at least one requested VM.
For instance, if $m$ nodes with availability $A_1$, $A_2$,\ldots, $A_m$ are used for hosting $k$ VMs ($m \leq k$), then the availability (denoted by $A_p$) of this VM placement is:
\begin{equation}
A_p=A_1 \cdot A_2 \cdots A_m  \label{Eq:AvbSimple}
\end{equation} 
Eq.~(\ref{Eq:AvbSimple}) indicates that since $k$ VMs are requested in total, the availability should take into account the probability that all these $k$ VMs are operational.

In the protected placement case, there exist one or more VMs that can be placed on at most $H$ nodes. Therefore, we regard that a protected placement $P$ is composed of (maximum) $H$ single placements. Within each single placement, the communication requirements between VM pairs should be satisfied. For the ease of clarification, we further term each of the $H$ single placements in the protected placement as placement group $p_i$, which means the ``$i-$th'' placing $k$ VMs on $m_i$ nodes, where $1 \leq i \leq H$ and $1 \leq m_i \leq k$. We regard $p_1$ as the primary placement group. We make no difference between the single placement and the placement group. Since different placement groups may place one or more VMs on the same node, we distinguish the protected placement as two cases, namely (1) \emph{fully protected placement}, for each VM $v \in V$, $v$ is placed by each group $p_i$ ($1 \leq i \leq H$) at $H$ different nodes, and (2) \emph{partially protected placement}, $\exists v \in V$, such that $v$ is placed on less than $H$ nodes, i.e., two or more placement groups place $v$ on the same node. 

In the fully protected placement case, the availability can be calculated as: 

\begin{flalign}
&A^{F}_{PD}=1-\prod_{i=1}^{H} (1-A_{p_{i}}) =\sum_{i=1}^{H}A_{p_{i}}-\sum_{0<i<j\leq H} A_{p_{i}} \cdot A_{p_{j}} \nonumber  \\
&+\sum_{ 0<i<j<u\leq H}A_{p_{i}} \cdot A_{p_{j}} \cdot A_{p_{u}}+\cdot\cdot\cdot+(-1)^{H-1}\prod_{i=1}^{H}A_{p_{i}} \label{Eq:Avb2F}
\end{flalign}
where $A_{p_i}=\prod_{n \in m_i} A_n$ denotes the availability of a single VM placement according to Eq.~(\ref{Eq:AvbSimple}). Eq.~(\ref{Eq:Avb2F}) reflects that the availability of $H$ placement groups is equivalent to the probability that at least one single placement (a set of $k$ VMs) is operational in the service-life time. 

\begin{figure}[tbh]
\centering
\subfloat[Without SRNG]{
\includegraphics[trim=5mm 5mm 5mm 5mm,clip=true,width=0.25\textwidth]{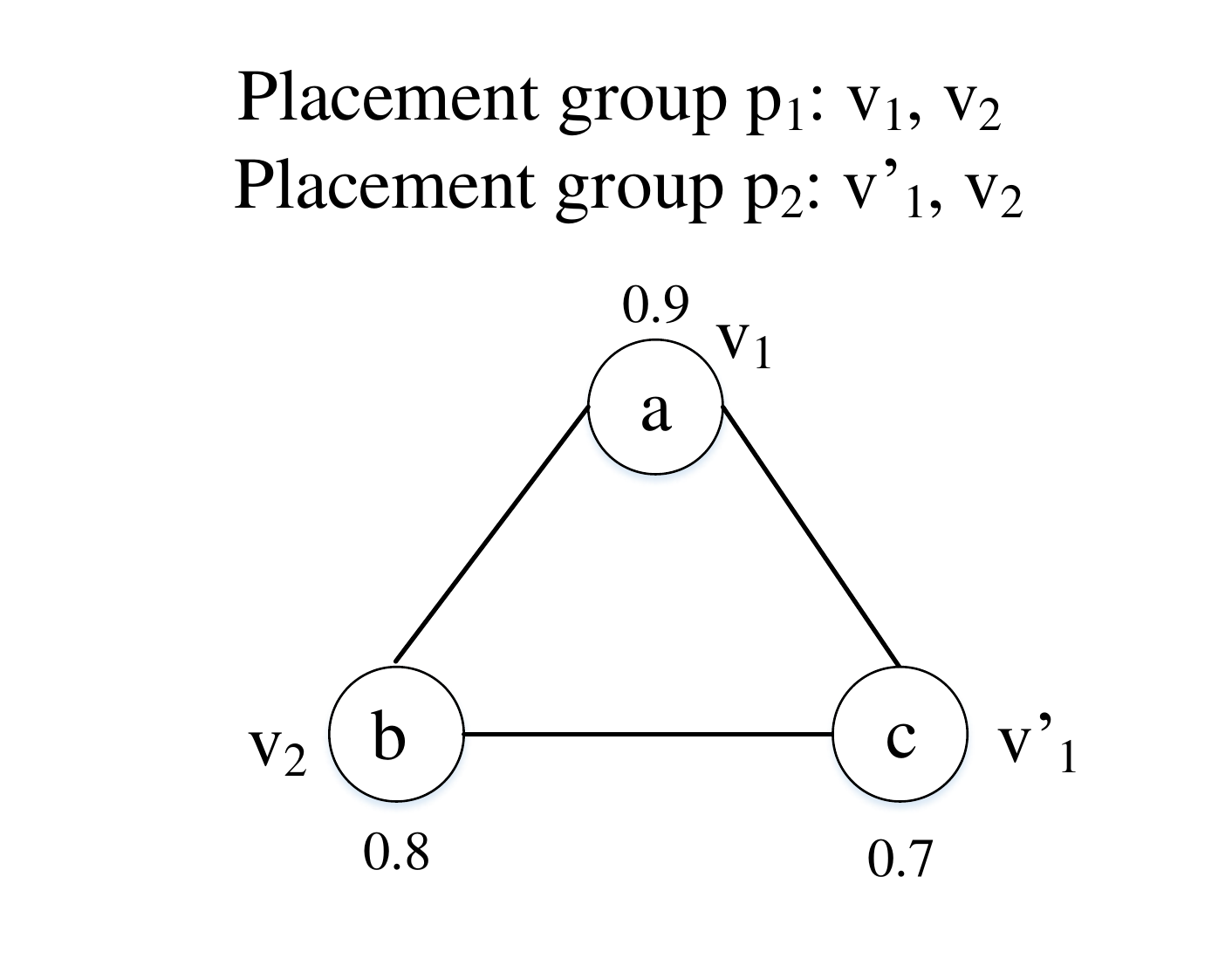}\label{Fig:DPRea}}
\subfloat[With SRNG]{
\includegraphics[trim=5mm 5mm 5mm 5mm,clip=true,width=0.25\textwidth]{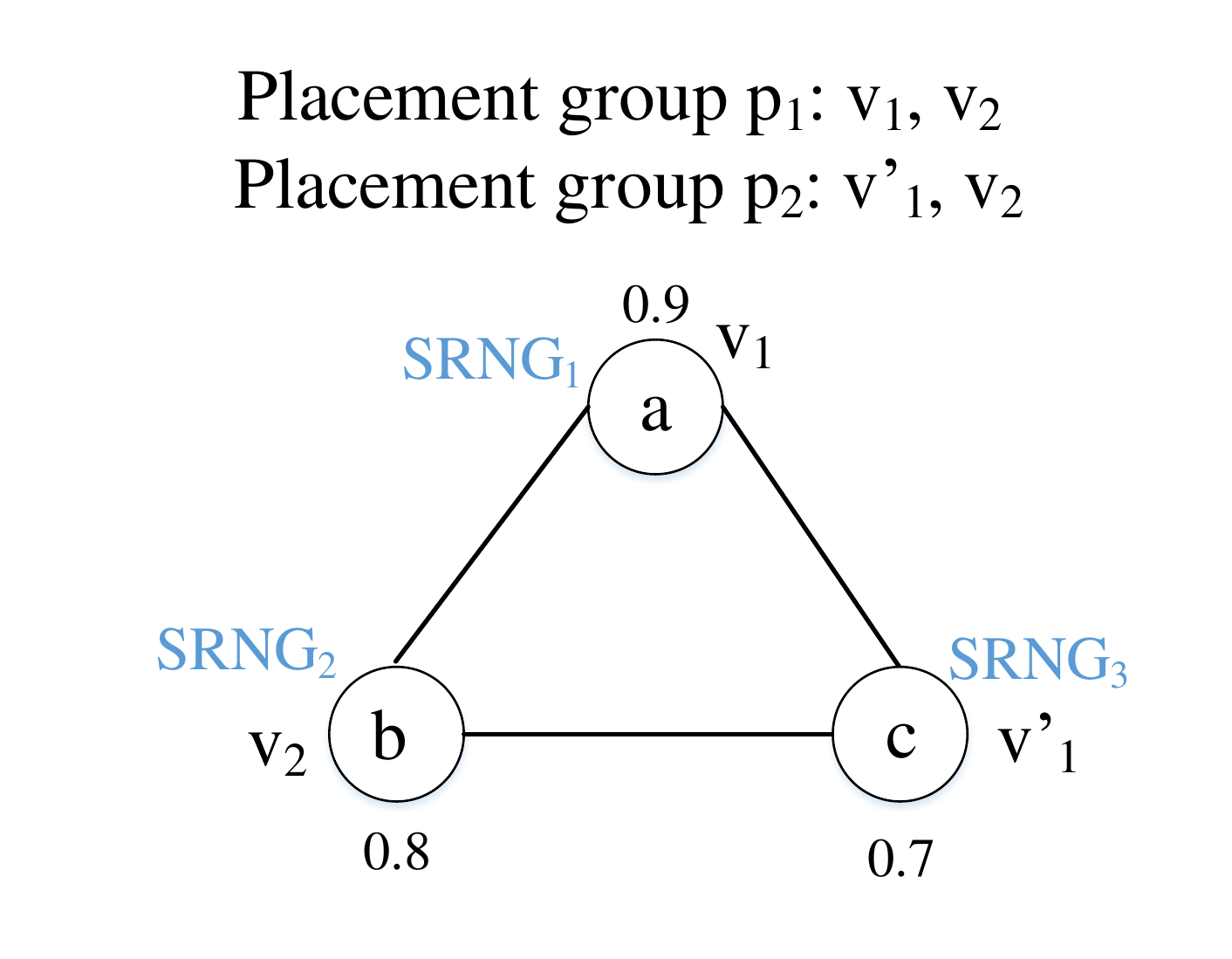}\label{Fig:SRNG}}
\caption{Partially VM placement availability calculation.}%
\label{Fig:AvbPlacement}%
\end{figure}

In the partially protected placement case, if one VM is placed on less than $H$ nodes, we could regard that this VM is jointly placed by more than one placement group. For example, in Fig.~\ref{Fig:AvbPlacement}\subref{Fig:DPRea}, each node is associated with its own availability value and we need to place two VMs ($v_1$ and $v_2$) on it. We set $H_1=2$ and $H_2=1$ for simplicity. We assume that placement group $p_1$ places $v_1$ on node $a$, and placement group $p_2$ places $v_1$'s replica (denoted by $v'_1$) on node $c$. On the other hand, $v_2$ is only placed on one node. Therefore, we can regard that $p_1$ and $p_2$ jointly place $v_2$ on node $b$. 

However, we cannot directly apply Eq.~(\ref{Eq:Avb2F}) to calculate its availability, since the availabilities of nodes which hold ``shared'' VMs will be counted more than once. 
To amend this, we use a new operator $\circ$\footnote{As in \cite{RNDM, Yang15} for the partially link-disjoint paths connection availability.}. Suppose there are $m$ different nodes $n_1, n_2, \ldots, n_m$ with availabilities $A_1, A_2, \ldots, A_m$. For a node $n_x$ with availability $A_x$, $\circ$ can be defined as follows:

\begin{equation}
A_{1} \cdot A_{2} \cdot \cdot \cdot A_{m} \circ A_x=\left \{
\begin{array}[l]{l}
\prod_{i=1}^{m}A_{i}   \quad \qquad \text{if} \quad \exists n_{i}=n_x \\
\prod_{i=1}^{m}A_{i} \cdot A_x \quad \text{ otherwise} 
\end{array}  \label{Eq: DefCirc}
\right.
\end{equation}

Let $\coprod$ denote consecutive $\circ$ operations of the different sets, then the availability (represented by $A^{H}_{PD}$) of $H$ partially placement groups can now be represented as:

\begin{align}
&A^{H}_{PD}=1-\coprod_{i=1}^{H} (1-A_{p_{i}})  \nonumber \\
&=1- (1-A_{p_{1}})\circ (1-A_{p_{2}}) \circ \circ \circ (1-A_{p_{H}}) \label{Eq:KPDisjointAva} \\
&=\sum_{i=1}^{H}A_{p_{i}}-\sum_{0<i<j\leq H} A_{p_{i}} \circ A_{p_{j}}+ \nonumber  \\
&\sum_{ 0<i<j<u \leq H}A_{p_{i}} \circ A_{p_{j}} \circ A_{p_{u}}+\cdot\cdot\cdot+(-1)^{H-1}\coprod_{i=1}^{H}A_{p_{i}} \nonumber
\end{align}
where $A_{p_i}$ denotes the availability of placement group $p_i$ and can be calculated from Eq.~(\ref{Eq:AvbSimple}). Now, going back to the example of Fig.~\ref{Fig:AvbPlacement}\subref{Fig:DPRea}, when there are no communication requirements between the two requested VMs, the placement availability of $p_1$ and $p_2$ is equal to $1-(1-A_a \circ A_b) \circ (1-A_c \circ A_b)=A_a \circ A_b +A_c \circ A_b-A_a \circ \underline{A_b} \circ A_c \circ \underline{A_b}=A_a \cdot A_b +A_c \cdot A_b-A_a \cdot \underline{A_b} \cdot A_c=0.9 \cdot 0.8+0.7 \cdot 0.8-0.9 \cdot 0.8 \cdot 0.7=0.776$.

\begin{figure}[tbh]
\centering
\includegraphics[trim = 0mm 0mm 0mm 0mm,clip,width=0.39\textwidth]{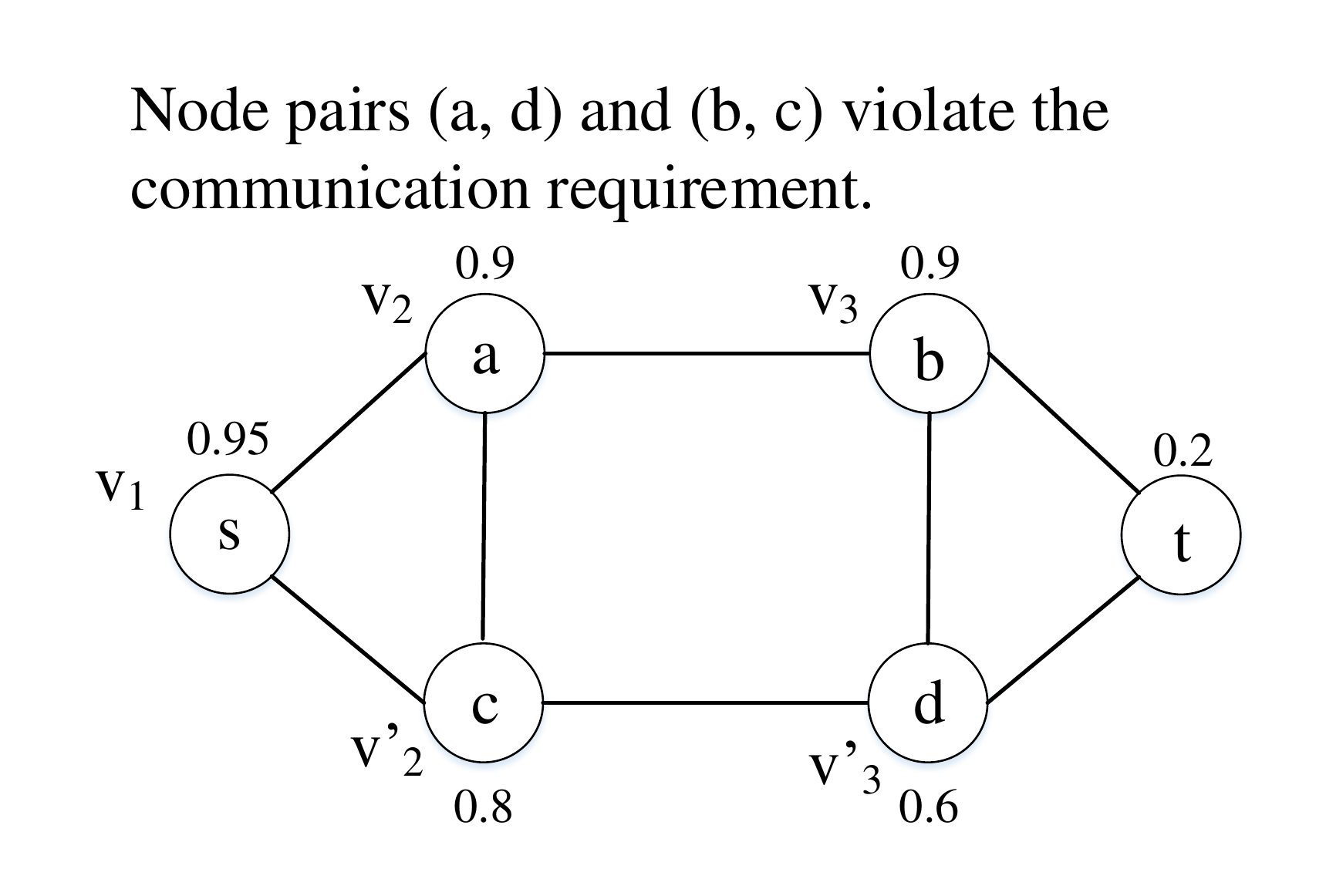}
\caption{An example of a partially VM placement availability calculation under communication requirement constraint.}%
\label{Fig:DPAvb}%
\end{figure}

In order to emphasize the importance of communication requirements between the VMs, we consider the example of Fig.~\ref{Fig:DPAvb}. For simplicity, it is assumed that each node can host at most one VM and its availability value is depicted in Fig.~\ref{Fig:DPAvb}. Moreover, we impose that node pairs $(a, d)$ and $(b,c)$ have communication delays bigger than the requested delay, i.e., these two node pairs do not satisfy the VM communication requirement.

As is shown in Fig.~\ref{Fig:DPAvb}, there are in total 4 possible placement groups: ($v_1$, $v_2$, $v_3$), ($v_1$, $v_2$, $v'_3$), ($v_1$, $v'_2$, $v_3$) and ($v_1$, $v'_2$, $v'_3$). However, neither ($v_1$, $v'_2$, $v_3$) nor ($v_1$, $v_2$, $v'_3$) can form a feasible placement group, because the communication delay is violated in either of these groups, and therefore, their availability cannot be taken into account.\footnote{As a side note, neglecting the communication requirement of VM pairs, the overall availability of the $4$ placement groups would have been: 
$1-(1-A_s \circ A_a \circ A_b) \circ (1-A_s \circ A_a \circ A_d)\circ (1-A_s \circ A_c \circ A_b)\circ (1-A_s \circ A_c \circ A_d)=0.89376$.} As a result, we only take into account the two placement groups ($v_1$, $v_2$, $v_3$) and ($v_1$, $v'_2$, $v'_3$). According to Eq.~(\ref{Eq:KPDisjointAva}), their overall availability is 
$1-(1-A_s \circ A_a \circ A_b) \circ (1-A_s \circ A_c \circ A_d)=A_s \circ A_a \circ A_b+ A_s  \circ A_c \circ A_c-\underline{A_s} \circ A_a \circ A_b \circ \underline{A_s} \circ A_c \circ A_d=A_s \cdot A_a \cdot A_b+A_s \cdot A_c \cdot A_c-\underline{A_s} \cdot A_a \cdot A_b \cdot A_c \cdot A_d=0.95 \cdot 0.9 \cdot 0.9+0.95 \cdot 0.9 \cdot 0.8-0.95 \cdot 0.9 \cdot 0.9 \cdot 0.8 =0.85614$.

\section{Shared-Risk Node Group} \label{Sec:SRNG}
In this section, we assume two types of failures/availabilities, namely Shared-Risk Node Group (SRNG) failures and single node failures/availabilities. A SRNG failure \cite{Datta08} reflects that a particular group of nodes will fail simultaneously, i.e., they have correlated failures. For example, in data center networks \cite{Kachris12}, servers are hosted by different racks, and within the same rack, servers are connected with a Top of Rack (ToR) switch. These ToR switches are further inter-connected through aggregate switches in a tree-like topology. Similarly, these aggregate switches are further connected with several core switches in the upper layer. In this context, the servers which are hosted by the same rack will fail simultaneously, if the rack they belong to fails. Similarly, the failure of a ToR switch will cause the failures of all the racks which are connected with it, that will subsequently cause failures of the hosted servers. One rack failure then corresponds to one distinct SRNG event. One (server) node can belong to multiple SRNG events (e.g., rack failure or switch failure). We assume there are in total $g$ SRNG events, and the failure probability of SRNG event $i$ is represented by $\pi_i$. For the ease of elaboration, we let $\lambda_i=1-\pi_i$, which can represent the non-occurring probability of SRNG event $i$.
For each node $n \in \mathcal{N}$, we denote $R_n$ as the set of all the SRNG events it belongs to. The VM placement availability should incorporate the SRNG non-occurring probabilities as well as the node availabilities. As a result, if $m$ nodes with availability $A_1$, $A_2$,\ldots, $A_m$ are used for hosting $k$ VMs ($m \leq k$) by a single placement $p$, then the availability of this single placement $p$ can be calculated as follows:

\begin{align} \label{Eq:SRNG}
 \prod_{i;srng_i \cap p \neq \emptyset } \lambda_i  \cdot  \prod_{j=1}^{m}A_j
\end{align}     

Similarly, the protected VM placement (fully and partially) availability can be calculated by substituting Eq.~(\ref{Eq:SRNG}) with $A_{p_i}$ in Eq.~(\ref{Eq:KPDisjointAva}). It is worthwhile to mention that the operator $\circ$ in Eq.~(\ref{Eq:KPDisjointAva}) still holds for $\lambda$. For example, in Fig.~\ref{Fig:AvbPlacement}\subref{Fig:SRNG}, each node is assigned with a SRNG event and node availability value. Suppose $\lambda_1=0.999$, $\lambda_2=0.99$ and $\lambda_3=0.9$. Assume that placement $p_1$ places $v_1$ on $A$ and $v_2$ on $B$, and placement $p_2$ places $v_1$ on $C$ and $v_2$ on $B$. In this sense, both the SRNG event failure and node availability should only be counted once, especially for $SRNG_2$ and node $B$ (since they both belong to $p_1$ and $p_2$). Consequently, the VM placement availability of $p_1$ and $p_2$ is:

\small
\begin{align*}
 &1-( 1-\lambda_1 \circ \lambda_2 \circ A_a \circ A_b) \circ (1-\lambda_2 \circ \lambda_3 \circ A_b \circ A_c ) \\
 =&\lambda_1 \circ \lambda_2 \circ A_a \circ A_b+\lambda_2 \circ \lambda_3 \circ A_b \circ A_c-  \\   \nonumber
 &\lambda_1 \circ \underline{\lambda_2} \circ A_a \circ \underline{A_b} \circ \underline{\lambda_2} \circ \lambda_3 \circ \underline{A_b} \circ A_c \\ \nonumber
 =&\lambda_1 \cdot \lambda_2 \cdot A_a \cdot A_b+\lambda_2 \cdot \lambda_3 \cdot A_b \cdot A_c-\lambda_1 \cdot \underline{\lambda_2} \cdot A_a \cdot \underline{A_b} \cdot \lambda_3 \cdot A_c \\ \nonumber
=&0.999 \cdot 0.99 \cdot 0.9 \cdot 0.8+0.99 \cdot 0.9 \cdot 0.8 \cdot 0.7- \\ \nonumber
& 0.999 \cdot 0.99 \cdot 0.9 \cdot 0.8  \cdot 0.9 \cdot 0.7=0.762432264
\end{align*}

\normalsize

\section{Reliable Virtual Machine Placement} \label{Sec:ProbDef}

\subsection{Problem Definition} \label{Subsec:RVMP}

We denote by $\mathcal{N}$ the set of $N$ server nodes and by $\mathcal{L}$ the set of $L$ links between them. The server nodes in $\mathcal{N}$ form a complete graph\footnote{There is a possibility of all nodes to be connected, but the quality/goodness of these connections are determined by other factors (e.g., connection availability, communication delay).} ($L=\frac{N(N-1)}{2}$). Each node $N_j \in \mathcal{N}$ has a storage upper bound of $s_j$. For each link $(m,n) \in  \mathcal{L}$, a function $F(m,n,\eta, D)$ returns $1$ if $m$ and $n$ can have a connection availability of at least $\eta$ and communication delay of at most $D$, and $0$ otherwise. A request is denoted by $r(k,c,V, T, A, \delta)$, where $k$ indicates the requested number of VMs $V$ with demanding capacity $c_v$ ($v \in V$). $T$ and $A$ are $k \times k$ matrices, which specify the delay constraint and connection availability constraint between any two VMs, respectively, and $\delta$ is the requested VM placement availability. 

Formally, the Reliable VM Placement (RVMP) problem is defined as follows:
\begin{definition} \label{Def:RVMP}
For a request $r(k,c,V, T, A, \delta)$, the Reliable VM Placement (RVMP) problem is to place at most $H$ groups of $k$ VMs on a minimum number of nodes such that: 
\begin{itemize}
\item The VM placement availability is no less than $\delta$. 
\item Each node does not exceed its storage limit.
\item Any two VMs $i_1$ and $i_2$ under the same placement group have a communication delay no more than $T(i_1,i_2)$, and a connection availability no less than $A(i_1,i_2)$.
\end{itemize}
\end{definition}

In the RVMP problem, we assume that each VM can be placed at up to $H$ different nodes. Moreover, we only consider the server nodes and ignore some other nodes in the network (e.g., router nodes, switch nodes). In fact, the link between each node pair in the RVMP problem actually implies a (set of) path(s) which may traverse some other intermediate nodes. Finding reliable and delay-sensitive paths could be easier within a tree-like data center network, but this problem becomes harder when the node pairs are located in different data centers (a more general network).  
As we proved in \cite{RNDM, Yang15}, the problem of finding $w \geq 2$ link-disjoint paths for which the connection availability is no less than a given value is already NP-hard and cannot be approximated to an arbitrary degree. In this sense, jointly considering the RVMP problem and reliable routing problem will make it even harder to solve. Therefore, we assume $F(m,n,\eta, D)$ is precalculated by using the algorithms proposed in Section~\ref{Sec:Routing}, where we will address how to find reliable and delay-sensitive paths by taking link availability and link delay into account. 





\begin{theorem}
The RVMP problem is NP-hard.
\end{theorem}

\begin{IEEEproof}
Let us first introduce the NP-hard Bin-Packing problem \cite{Garey79}: Given $n$ items with sizes $e_1$, $e_2$, \ldots , $e_n$, and a set of $m$ bins with capacity $c_1$, $c_2$, \ldots , $c_m$, the Bin-Packing problem is to pack all the items into minimized number of bins without violating the bin capacity size. 
If we assume that for each node pair $(m,n)$, $F(m,n,\eta,D)=1$ for any $\eta$ and $D$ and all the nodes have availability $1$, then the RVMP problem for $H=1$ is equivalent to the Bin-Packing problem, which is NP-hard. 
Next, let us analyze its complexity when the objective of minimizing the number of used nodes is not considered. 

\begin{itemize}
\item Each node has unlimited storage: In this case, each set of $k$ VMs can be placed on one node and we need to find $H$ nodes in the network to store each set of $k$ VMs. This can be solved in $N \choose H $ searching when $N>H$ or using $N$ nodes to host $N$ groups of $k$ VMs when $N \leq H$, which is polynomial time solvable.

\item Each node has limited storage: Assume $H=1$ and $A_n=1$, $\forall n \in \mathcal{N}$. Moreover, assume a certain $D$ value and that $(m,n) \in \mathcal{L} $, $F(m,n,\eta,D)$ remains the same for any $\eta$. That is, the link between each node pair is only assigned with a delay value (connection availability is not taken into account). Under this assumption, Alicherry and Lakshman \cite{Alicherry13} have proved that the RVMP problem can be reduced to the 3SAT problem, and cannot be approximated to an arbitrary degree. 
\end{itemize}

\end{IEEEproof}
The RVMP problem with SRNG failures is also NP-hard and cannot be approximated to an arbitrary degree, when we assume that each node is associated with one distinct SRNG event and all the node availabilities are assumed to be $1$. In the following, we will devise both an exact solution and a heuristic to solve the RVMP problem. 

\subsection{Exact Solution} \label{Subsec:ExactAlg}
In this subsection, we propose an exact Integer Nonlinear Program (INLP) to solve the RVMP problem. We first solve the RVMP problem without SRNG failures and start by explaining the necessary notations and variables:

\textbf{INLP notations:}

$r(k,c,V,T,A, \delta)$: A VM placement request $r$ as specified in Section~\ref{Subsec:RVMP}.  

$\mathcal{N}, \mathcal{L}$: set of $N$ nodes and set of $L$ links, respectively.

$H$: The maximum number of times for one VM to be placed in the network. 

$F(m,n,\eta, D)$ Returns $1$ if a connection exists between $m$ and $n$ such that connection availability is at least $\eta$ and communication delay value is at most $D$, and $0$ otherwise.

$\lambda_i^n$: The non-occurring probability of the $i$-th SRNG if node $n$ belongs to it, and $1$ otherwise.

\textbf{INLP variable:}

$P_{vn}^h$: a binary variable and it is equal to $1$ if VM $v$ is placed on node $n$ by placement group $h$, and $0$ otherwise, where $v \in  V$, $n \in \mathcal{N}$ and $1 \leq h \leq H$.

\textbf{Objective:}
\begin{equation} \label{Eq:Obj}
\min  \sum_{n \in \mathcal{N}}  \left( \max_{1 \leq h \leq H,v \in V}  P_{vn}^h \right)
\end{equation}

\textbf{Placement constraint:}

\begin{flalign}
\sum_{n \in \mathcal{N}, 1 \leq h \leq H} P^h_{vn} \geq 1 ~~~\forall v \in V
\label{Eq: PlaceConstraint}
\end{flalign}

\textbf{Storage constraint:}

\begin{flalign}
\sum_{v \in V} \left( \max_{h=1}^H P_{vn}^h \right) \cdot c_v   \leq s_n ~~~~~ \forall n \in \mathcal{N}
\label{Eq: BandConstraint}
\end{flalign}

\textbf{Delay and connection availability constraint:}

\begin{flalign}
F(m,n,A(m,n),T(m,n)) \cdot P^h_{am} \cdot P^h_{bn}=1  ~~~\nonumber \\  \forall 1 \leq h \leq H, (m,n) \in \mathcal{L},       
1 \leq a, b \leq k: a \neq b 
\label{Eq: DelayConstraint}
\end{flalign}

\textbf{VM placement availability constraint:}

\footnotesize
\begin{flalign}
&\sum\limits_{h=1}^{H}\prod\limits_{n \in \mathcal{N}} \min \limits_{v \in V} \left(
1-P^h_{vn}+P^h_{vn}A_{n}\right)-   \notag \\
& \sum\limits_{1 \leq h<u\leq H}\prod\limits_{n \in \mathcal{N}}\min
\left( \min \limits_{v \in V} (1-P_{vn}^h+P_{vn}^h A_{n}),\min \limits_{v \in V} (1-P_{vn}^u+P_{vn}^u A_{n})\right)  \notag \\
& +\cdot \cdot \cdot +(-1)^{H-1}\left( \prod\limits_{n \in \mathcal{N}}\min_{1\leq h\leq
H}( \min \limits_{v \in V} (1-P_{vn}^h+P_{vn}^h A_{n}))\right)\geq \delta
\label{Eq: AvbConstraint}
\end{flalign}

\normalsize
Eq.~(\ref{Eq:Obj}) minimizes the number of total used nodes. For instance, we first calculate the maximum value of $P^h_{vn}$ for node $n \in \mathcal{N}$, and as long as $P^h_{vn}=1$ for some $1 \leq h\leq H$ and $v \in V$, it means that node $n$ is in use to host VM(s). After that, we take the sum of $ \max_{1 \leq h \leq H, v \in V}  P_{vn}^h$ for all the nodes in $\mathcal{N}$ and try to minimize this value.
Eq.~(\ref{Eq: PlaceConstraint}) ensures that each one of $k$ requested VMs must be placed in the network. 
Eq.~(\ref{Eq: BandConstraint}) ensures that each node does not exceed its storage limit when VMs are placed on it. 
Eq.~(\ref{Eq: DelayConstraint}) makes sure that the specified delay constraint and connection availability of any two VMs under the same placement group are not violated.
Eq.~(\ref{Eq: AvbConstraint}) ensures that the VM placement availability constraint is obeyed, according to Eq.~(\ref{Eq:KPDisjointAva}). We note that Eq. (\ref{Eq: AvbConstraint}) can simultaneously calculate the availability of the fully protected placement, partially protected placement, and single placement. For instance, when $H=2$, Eq. (\ref{Eq: AvbConstraint}) becomes:

\small
\begin{flalign}
&\prod \limits_{n\in \mathcal{N}} \min \limits_{v \in V} (1-P_{vn}^{1}+P_{vn}^{1}A_{n})+
\prod\limits_{n\in \mathcal{N}} \min \limits_{v \in V} (1-P_{vn}^{2}+P_{vn}^{2}A_{n})-  \nonumber \\
&\prod \limits_{n\in \mathcal{N}}\min(\min \limits_{v \in V}(1-P_{vn}^{1}+P_{vn}^{1}A_{n}), \min \limits_{v \in V} (1-P_{vn}^{2}+P_{vn}^{2}A_{n})) \geq \delta \label{Eq: 2AvbConstraint}
\end{flalign}

\normalsize
When $P_{vn}^{1}=P_{vn}^{2}$ for all $ n \in \mathcal{N}$, Eq. (\ref{Eq: 2AvbConstraint}) becomes
\begin{equation}
\prod \limits_{n \in \mathcal{N}} \min \limits_{v \in V} (1-P_{vn}^{1}+P_{vn}^{1}A_{n}) \geq \delta \nonumber
\end{equation}
which is the VM placement availability constraint for the single placement. 


To solve the RVMP problem with SRNG failures, we need to rewrite Eq.~(\ref{Eq: AvbConstraint}) in Eq.~(\ref{Eq: SRNGAvbConstraint}) and keep the objective and all other constraints the same (Eq.~(\ref{Eq:Obj})-Eq.~(\ref{Eq: DelayConstraint})). 

Although inefficient in practice when the problem size is large, the INLP is useful for comparison purposes and demonstrates how accurate the heuristics are. This is shown in Figures~\ref{Fig:SimkH2} and \ref{Fig:SimkH3} (for smaller size problems). 
\begin{figure*}[!h]
\scriptsize
\begin{flalign}
&\sum\limits_{h=1}^{H}  \prod\limits_{n \in \mathcal{N}}  \min \limits_{v \in V} \left(
1-P^h_{vn}+P^h_{vn}A_{n}\right) \cdot \prod \limits_{1 \leq i \leq g} \min_{v \in V, n \in \mathcal{N}} \left(1-P^h_{vn}+P^h_{vn} \lambda_i^n  \right)   -   \notag \\
& \sum\limits_{1 \leq h<u\leq H}\prod\limits_{n \in \mathcal{N}}\min
\left( \min \limits_{v \in V} (1-P_{vn}^h+P_{vn}^h A_{n}),\min \limits_{v \in V} (1-P_{vn}^h+P_{vn}^h A_{n})\right) \cdot \prod \limits_{1 \leq i \leq g} \min ( \min_{v \in V, n \in \mathcal{N}} \left(1-P^u_{vn}+P^u_{vn} \lambda_i^n  \right), \min_{v \in V, n \in \mathcal{N}} \left(1-P^u_{vn}+P^u_{vn} \lambda_i^n  \right)) \notag \\
& +\cdot \cdot \cdot +(-1)^{H-1}\left( \prod\limits_{n \in \mathcal{N}}\min_{1\leq h\leq
H}( \min \limits_{v \in V} (1-P_{vn}^h+P_{vn}^h A_{n}))\right) \cdot \prod \limits_{1 \leq i \leq g} \min_{ 1 \leq h \leq H} \left( \min_{v \in V, n \in \mathcal{N}}  \left(1-P^h_{vn}+P^h_{vn} \lambda_i^n \right) \right) \geq \delta
\label{Eq: SRNGAvbConstraint}
\end{flalign}
\end{figure*}

\normalsize
\subsection{Heuristic Algorithm}  \label{Subsec:Heu}

\begin{algorithm}[h]
\caption{DSR$(\mathcal{G}(\mathcal{N},\mathcal{L}), r(k,c,V,T,A, \delta), H, \alpha)$} \label{Alg: DSR}
  $VP[h][v][n] \leftarrow 0$~$\forall  1 \leq h \leq H,  |v|=k, |n|= N$ \\
 \For{$h \leftarrow 1 $  \KwTo $H$ }
 {
   $VP[h] \leftarrow$ DSRPlace$(\mathcal{G}(\mathcal{N},\mathcal{L}), r(k, c, V, T, \delta), H, \alpha)$ \\
   $\mathcal{N} \leftarrow \mathcal{N} \backslash \mathcal{N}_x$, where $\mathcal{N}_x$ denotes a subset of the used nodes for already found placement groups.\\
   \If{$1- \coprod_{i=1}^H(1-A_{VP[i]}) \geq \delta$ }
   {
   		Call \footnotesize{PartiallyDSRPlace$(\mathcal{G}, VP[h][k][N], r, H)$}
   }
 }
 Return null 

\end{algorithm}

\begin{algorithm}[h]
\caption{DSRPlace$(\mathcal{G}(\mathcal{N},\mathcal{L}), r(k,c,V,T,A, \delta), H, \alpha)$} \label{Alg: DSRPlace}
\ForEach{$v_m$ in $V$~~$(1 \leq m \leq k)$}
{
   $v_x \leftarrow v_m$, $Q \leftarrow \emptyset $, $\mathcal{G}^m \leftarrow \mathcal{G}$, $P^m[V][N] \leftarrow 0$ \\
   \While{$Q.$Count $< k$}
   {
      Sort the nodes in $\mathcal{G}^m$ by their availabilities in the decreasing order $n_1$, $n_2$,\ldots ,$n_N$ \\
      Find one node $n_a$ with maximum availability to host $v_x$ without violating the delay and connection availability constraint with already placed VMs, such that $s_{n_a} \geq c_{v_x}$ \\
      \eIf{Step 5 succeeds}
      {
       $P^m[v_x][n_a] \leftarrow 1$, $s_{n_a} \leftarrow s_{n_a}-c_{v_x}$, $A_{n_a} \leftarrow 1 $, $Q$.Add($v_x$)\\
      }
      {
        Break\;
      }
      $\chi \leftarrow + \infty $\;
      \ForEach{$v_i$ in $V \backslash Q$}
      {
        \ForEach{$v_j$ in $Q$}
        {
           \If{$\chi> \frac{T(v_i, v_j) \cdot \alpha }{A(v_i, v_j)}  $}
           {
           	  $\chi \leftarrow \frac{T(v_i, v_j) \cdot \alpha }{A(v_i, v_j)}  $, $v_x \leftarrow v_i $ \\
           }
        }
      }
      
   }
  
}
 Return $P^m$ with the maximum availability.
\end{algorithm}

\begin{algorithm}[h]
\caption{PartiallyDSRPlace$(\mathcal{G}, VP,r, H)$}  \label{Alg: PartiallyDSR}
 Sort the nodes that host VMs in increasing order. \\
 Denote this set as $\mathcal{N}_y$.\\
 \ForEach{$n \in \mathcal{N}_y $  }
 {
   $VB \leftarrow VP$ \\
   $VB[h][v][n] \leftarrow 0 $ for $1 \leq h \leq H$ and $v \in V$ \\
		
   \ForEach{placement group $h=1...H$}
   {
	   Try to use its other used nodes to host VMs.\\
   }
   
    \If{$1- \coprod_{i=1}^H(1-A_{VB[i]}) \geq \delta$}
   {
     $VP \leftarrow VB$, $VB \leftarrow \emptyset$. 
   }

 }
 Return $VP$ 

\end{algorithm}

Our proposed heuristic to solve the RVMP problem, called the Delay-Sensitive and Reliable (DSR) placement algorithm, is shown in Algorithm~\ref{Alg: DSR}. Instead of placing VMs on nodes, the logic of DSR is to assign nodes to VMs until all the VMs are hosted by the nodes without violating VM delay and connection availability constraints. Since we want to use the least number of nodes to host VMs to satisfy the availability requirement, we gradually increase the amount of finding placement groups. In what follows, we explain each step of the heuristic algorithm, where the SRNG failures are first not considered. 

In Step $1$ of Algorithm~\ref{Alg: DSR}, we first initialize a binary variable $VP[h][v][n]$ representing whether VM $v \in V$ is hosted by node $n \in \mathcal{N}$ under the group $h$. After that, for placement group $h$, we call Algorithm~\ref{Alg: DSRPlace} to place VMs on nodes in Step $3$. The purpose of Step $4$ is to avoid different groups to have the same placement result. But this will only happen when a single node's free capacity is far greater than the VM demanding capacity. That is, all the VMs can be placed on the same node and its remaining free capacity is still large enough so that another set of $k$ VMs can be placed on it. In Step $5$, we calculate the availability of $VP[1]$, \ldots, $VP[h]$. If availability value of these $h$ placement groups is no less than $\delta$, we call Algorithm~\ref{Alg: PartiallyDSR} trying to return a partially VM placement solution in order to further reduce the number of used nodes. In Algorithm~\ref{Alg: PartiallyDSR}, for each node $n \in \mathcal{N}_y$, where $\mathcal{N}_y$ stores the nodes in the increasing order by their availabilities, we first clear all the VMs resident on $n$. For each placement group $p$, we try to use its other used nodes to host the VMs that are originally placed by it on $n$. For simplicity, we apply a greedy approach: for each one (say $n_u$) of used nodes by placement group $h$, we let $n_u$ host the VMs which are originally placed on $n$ by $p$ as many as possible. 
After that, we calculate whether the whole availability still satisfies $\delta$. If so, we assign this partially placement solution to $VP$.  
Next, we will explain the details of Algorithm~\ref{Alg: DSRPlace}, which is to find a single placement.

In Step $1$ of Algorithm~\ref{Alg: DSRPlace} we start with each $v_m \in V$, and assign it to $v_x$ in Step $2$. We use a queue $Q$ to store the VMs already placed, and initially it is set to empty.  
Besides, we also define variable $P^m[v][n]$ to indicate whether VM $v \in V$ is hosted by node $n \in \mathcal{N}$ corresponding to the placement group starting with VM $v_m$. 
As long as $Q$'s count is less than $k$, Step $4$-Step $9$ are going to assign nodes to host unassigned VMs. Step $5$ tries to find a node $n_a$ with maximum availability whose capacity should be at least $c(v_x)$. Moreover, if $v_x$ is placed on $n_a$, it should not violate the delay and connection availability constraints with already hosted VMs. If it succeeds, in Step $7$, the capacity of $n_a$ is reduced by $c_{v_x}$, the availability of $n_a$ is changed to $1$, and $v_x$ is added to $Q$. The reason to change a node's availability is that if some nodes have been used to host the existing VM(s), then the availability for these nodes to host other (unassigned) VMs is $1$. So we need to change its ``availability'' after each iteration of covering VM(s). If such a node cannot be found in Step $5$, this indicates that not all the VMs are covered and we consider this placement group should ``jointly'' place uncovered VMs with one of $H-1$ placement groups found in Algorithm~\ref{Alg: DSR}. The algorithm then breaks in Step $9$. Following that, Step $10$-Step $14$ search for an unsigned VM, which has the smallest value of $\frac{T(v_i, v_j) \cdot \alpha }{A(v_i, v_j)}$ to the VMs already placed, where $\alpha$ is a user given value. By doing this, we want to find an unsigned VM that has ``smaller'' path delay and ``greater'' connection availability constraints with already placed VMs, and assigns it to $v_x$. The motivation here is that we always first place the VM which has a more critical requirement in terms of path delay and connection availability.  
When $Q$'s count is equal to $k$, it indicates that all the VMs have been hosted, which means we get a ``complete'' placement group. Finally, in Step $15$, the algorithm returns a placement group with the biggest availability from $k$ already determined single placements.  

To solve the RVMP problem with SRNG failures, Alg.~\ref{Alg: DSR}-\ref{Alg: PartiallyDSR} remain the same except:

\begin{itemize}
\item In Step $4$ of Algorithm \ref{Alg: DSRPlace} and Step $1$ of Algorithm \ref{Alg: PartiallyDSR}, we sort the nodes in $\mathcal{G}^m$ by the product of their availabilities and non-occurring probabilities of all their belonging SRNG events in a decreasing order and an increasing order, respectively.

\item In Step $5$ of Algorithm \ref{Alg: DSRPlace}, we will find one node $n_a$ with maximum node availability multiplied by the non-occurring probabilities of SRNG events set $R_{n_a} \backslash R_x$, where $R_x$ denotes the set of SRNG events that $n_a$ belongs to but has already been considered/counted by the other nodes from Step $4$ to Step $14$. The reason is that one unique SRNG event can only occur once, so we cannot calculate its value under the same placement group more than once.  

\end{itemize}

The time complexity of Algorithm~\ref{Alg: DSRPlace} can be calculated like this: There are $k$ VMs in total in Step $1$, and Step $3$ has also $k$ iterations. Sorting algorithm for instance like insertion sort in Step $4$ takes $O(N\log (N))$ time, and Step $5$ has a complexity of $O(N)$.  
Step $9$-Step $13$ consume at most $O(k^2)$ time. Therefore, the whole complexity of Algorithm~\ref{Alg: DSRPlace} is $O(k^2 (N \log (N)+k^2))$. 
In Algorithm~\ref{Alg: PartiallyDSR}, Step $1$ consumes $O(N \log(N))$ time via insertion sort and Steps $2$-$8$ consume $O(N^2H)$ time, leading to a whole complexity of $O(N(\log N+NH))$. Consequently, the whole time complexity of Algorithm~\ref{Alg: DSR} is $O(k^2H(N \log (N)+k^2))$, since it calls at most $H$ times of Algorithm~\ref{Alg: DSRPlace}.

\subsection{Simulations} \label{Sec:Simulation}

The simulations are run on a desktop PC with $2.7$ GHz and $8$ GB memory. We use an Intel(R)Core(TM)i5-4310M CPU 2.70GHz x64-based processor in our simulations.  We use IBM ILOG CPLEX $12.6$ to implement the proposed INLP. All the heuristics are implemented by C\# and compiled on Visual Studio 2015 (using .NET Framework 4.5).

We set $\alpha=1$ for our heuristic DSR. We compare our exact INLP and heuristic DSR with two heuristics, namely (1) Greedy Placement (GP) and (2) Random Placement (RP). These 2 algorithms follow the similar routine with Algorithm~\ref{Alg: DSR}, except: (1) in Step $6$, they directly return the placement result if its availability is satisfied, instead of checking partially placement solution, and (2) they call different heuristics in Step $3$ (different from Algorithm~\ref{Alg: DSRPlace}), which we specify as follows:
\begin{itemize}
\item GP (or RP): It first selects a node with greatest availability (or randomly selects a node) and places as many VMs as possible on it under its storage limit. It then selects the second largest availability node (or randomly selects the second node) and places as many of the remaining VMs as possible, which should also satisfy the delay and connection availability constraints with the VMs already placed. This procedure continues until all the VMs are placed or all the nodes have been iterated.


\end{itemize}
In the following, we first test the algorithms for the RVMP problem without SRNG failures for both $16$-node and $100$-node networks, and then evaluate them for the RVMP problem with SRNG failures for a $100$-node network.

\subsubsection{$16$-node network without SRNG failures} \label{Sec:SimuSmall}

\begin{figure*}[!h]
\centering
\subfloat[Acceptance Ratio (AR)]{
\includegraphics[trim=25mm 90mm 25mm 88mm,clip=true,width=0.33\textwidth]{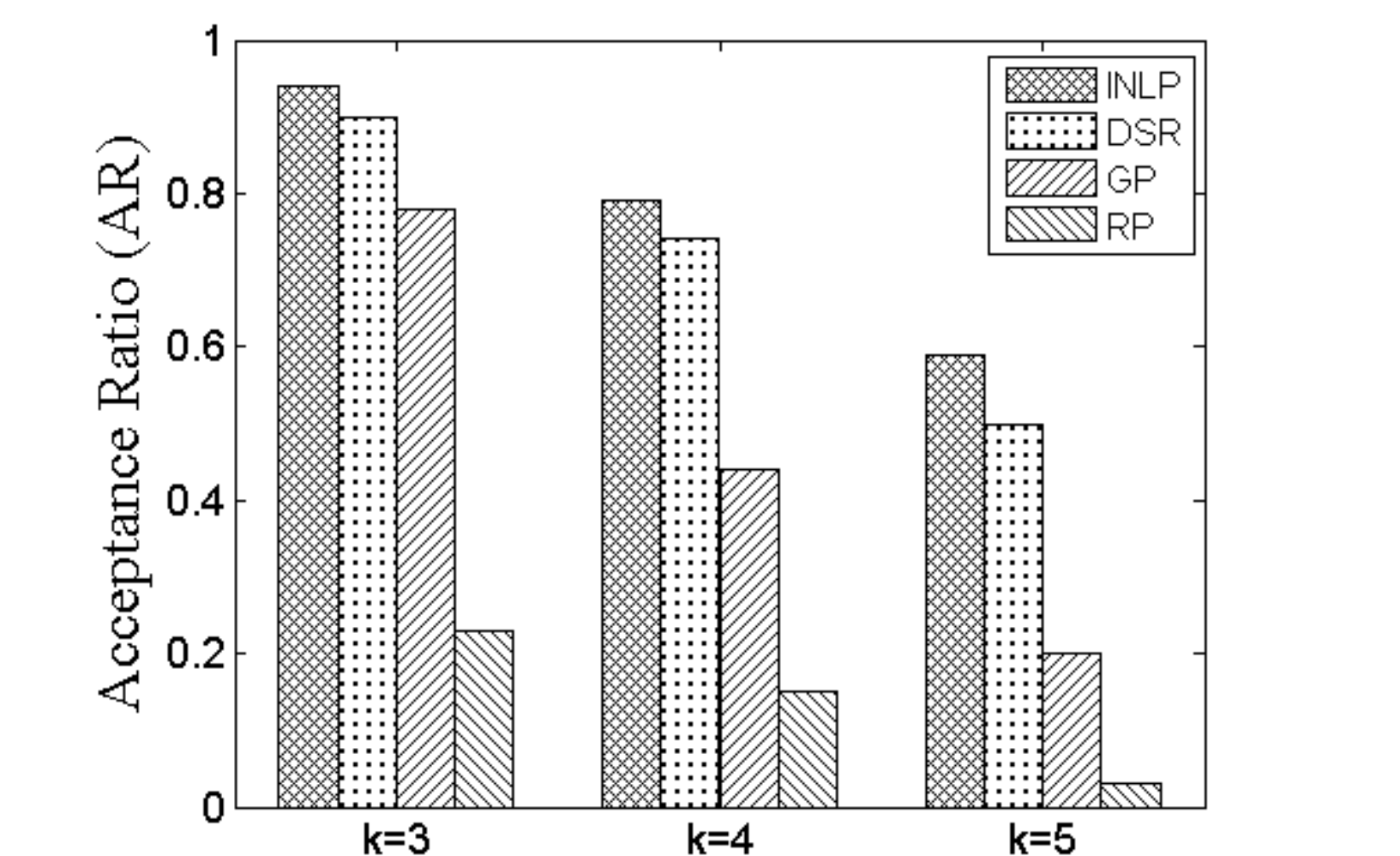}\label{Fig:ARH2}}
\subfloat[Average Number of Used Nodes]{
\includegraphics[trim=25mm 90mm 25mm 88mm,clip=true,width=0.33\textwidth]{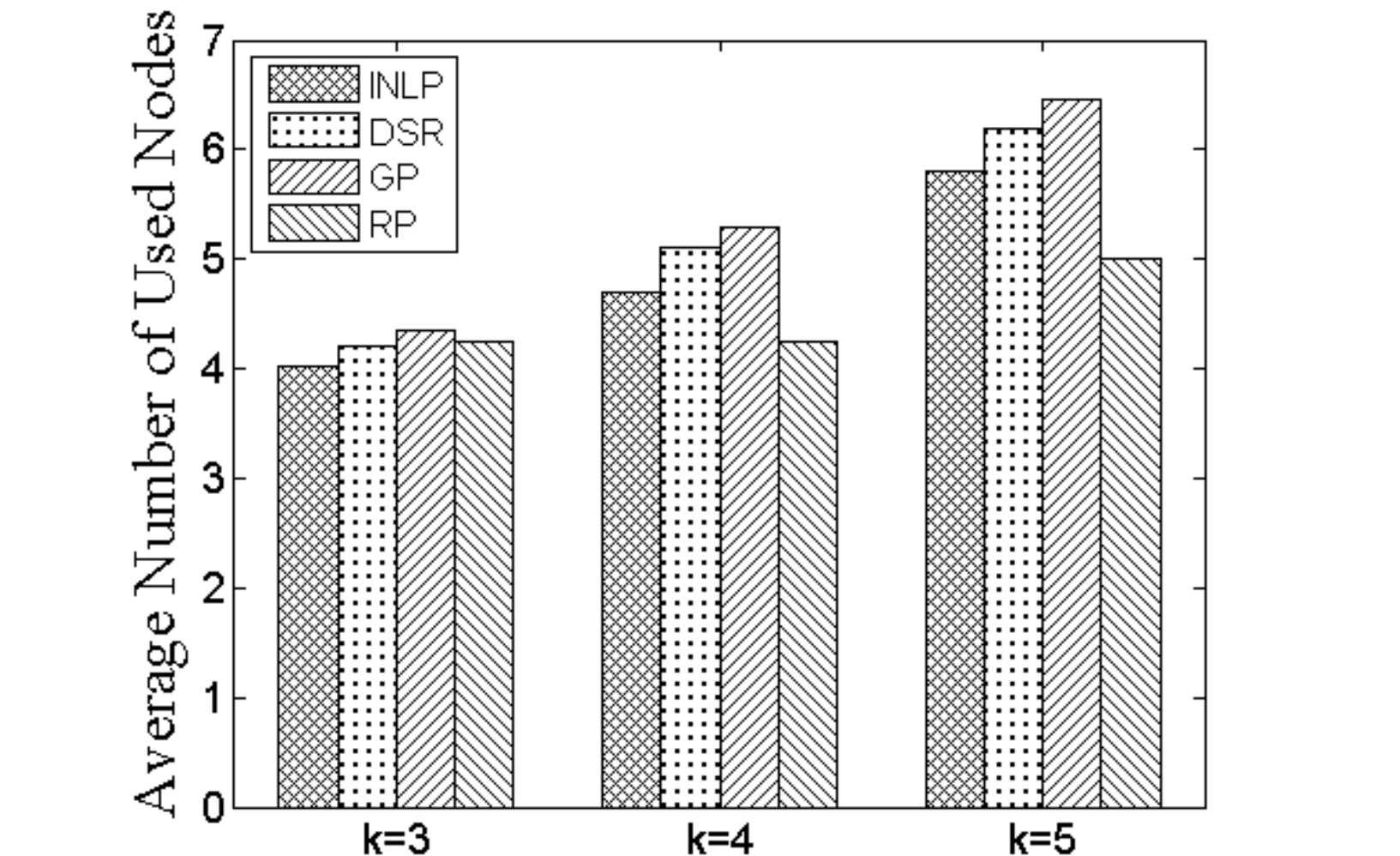}\label{Fig:ANUNH2}}
\subfloat[Running Time]{
\includegraphics[trim=25mm 90mm 25mm 88mm,clip=true,width=0.33\textwidth]{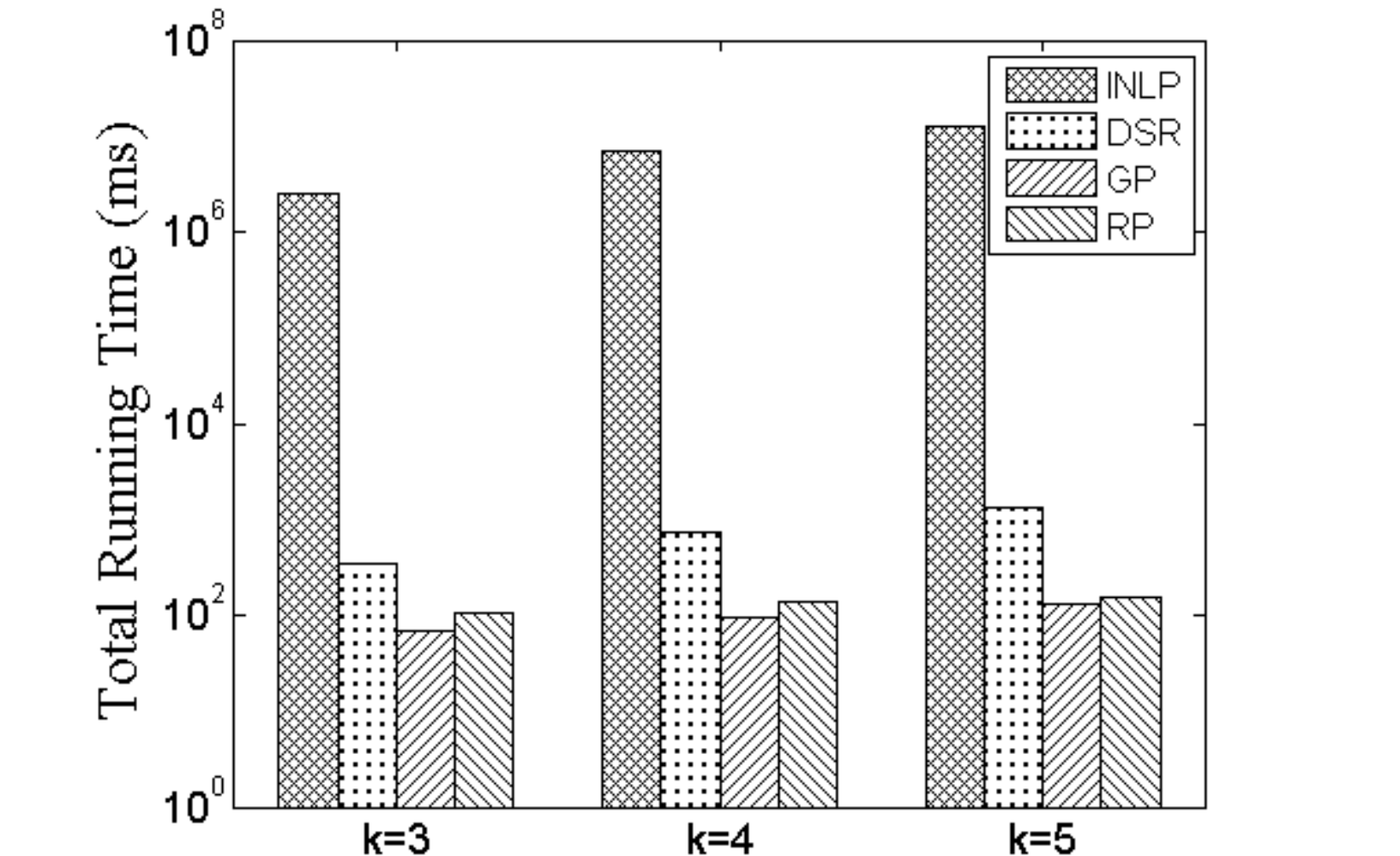}\label{Fig:TimeH2}}
\caption{Simulation results over 100 requests when $H=2$ for $16$-node network: (a) Acceptance Ratio (AR), (b) Average Number of Used Nodes (ANUN) and (c) Running Time.}%
\label{Fig:SimkH2}%
\end{figure*}
\begin{figure*}[!h]
\centering
\subfloat[Acceptance Ratio (AR)]{
\includegraphics[trim=25mm 90mm 25mm 88mm,clip=true,width=0.33\textwidth]{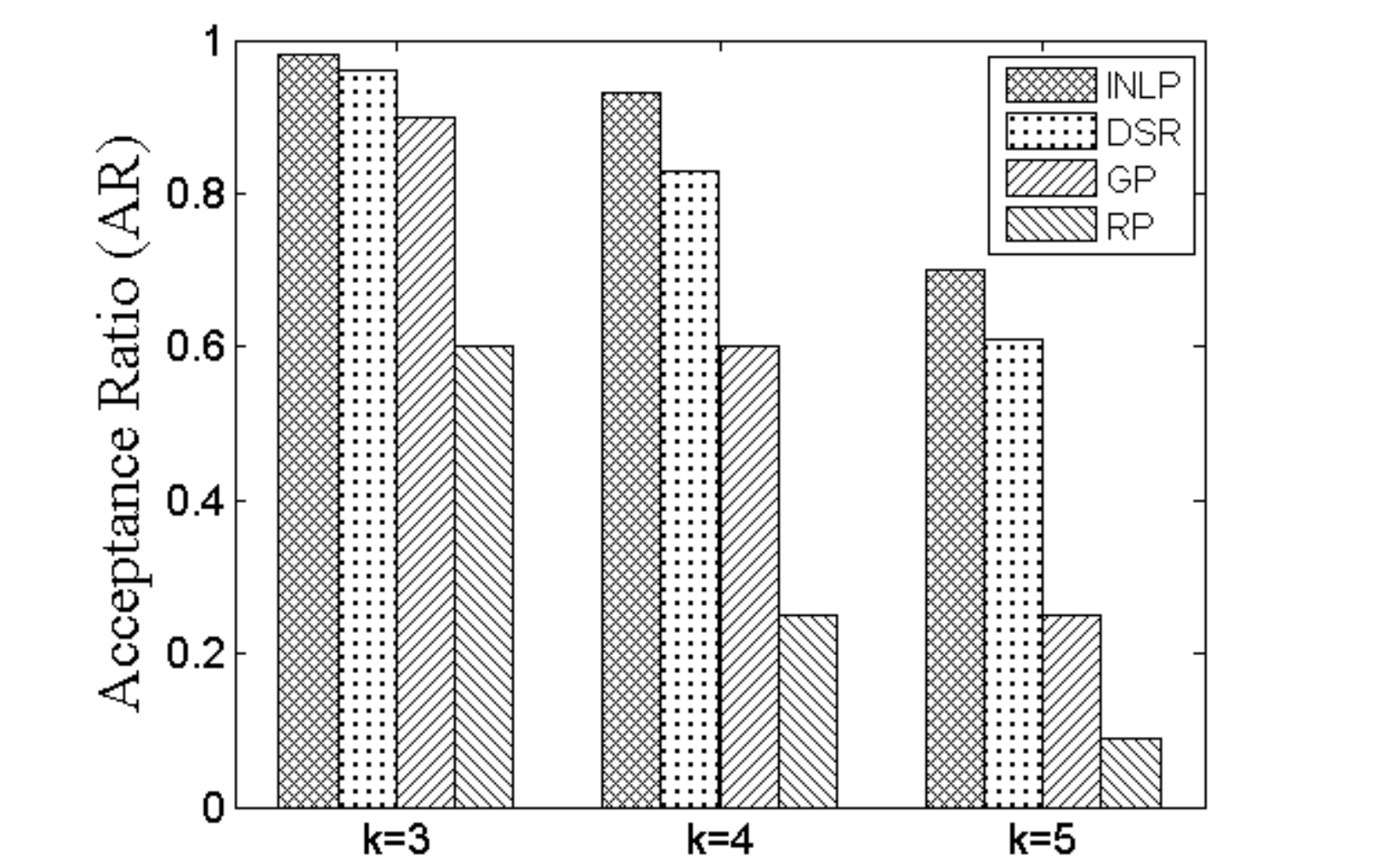}\label{Fig:ARH3}}
\subfloat[Average Number of Used Nodes]{
\includegraphics[trim=25mm 90mm 25mm 88mm,clip=true,width=0.33\textwidth]{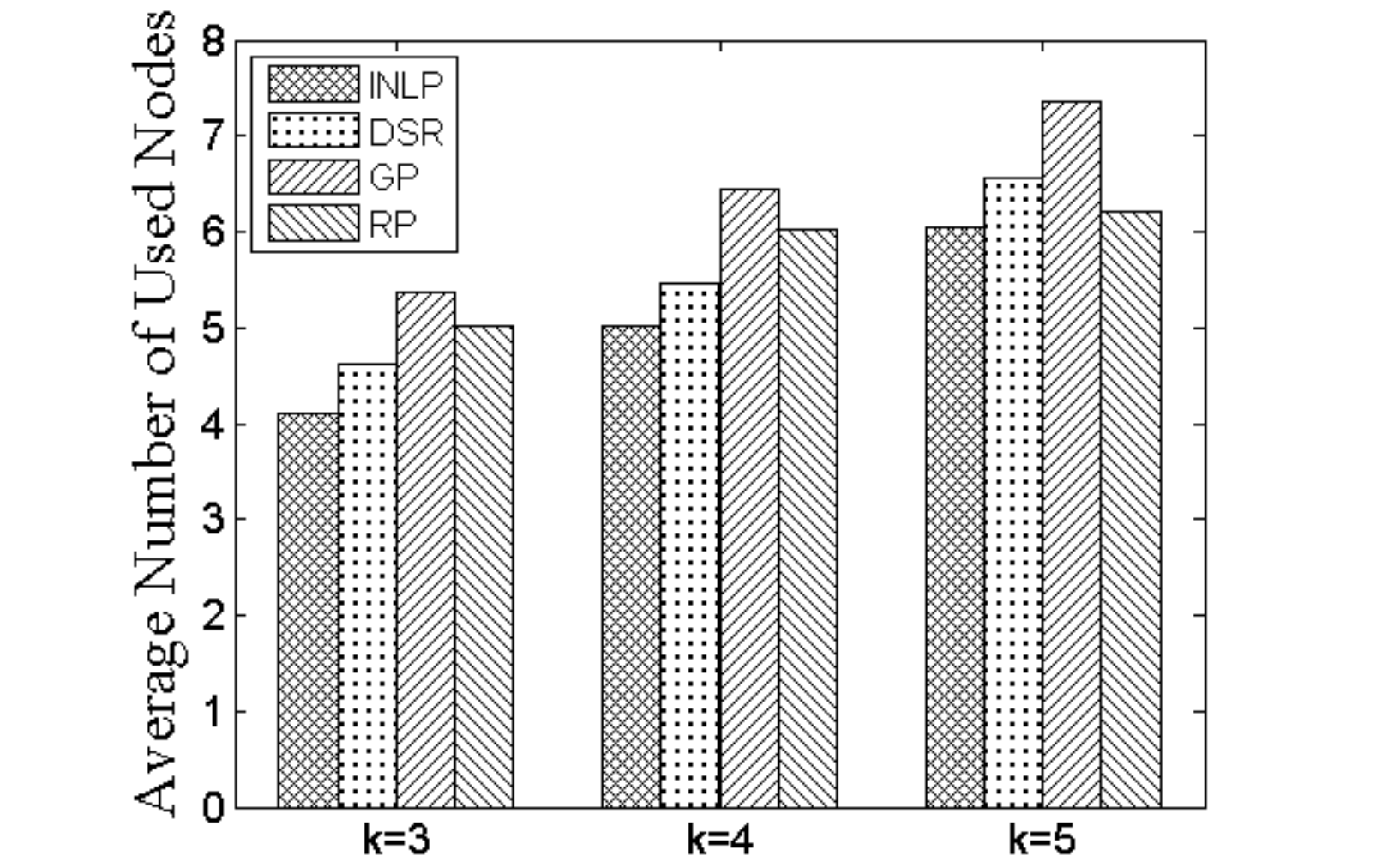}\label{Fig:ANUNH3}}
\subfloat[Running Time]{
\includegraphics[trim=25mm 90mm 25mm 88mm,clip=true,width=0.33\textwidth]{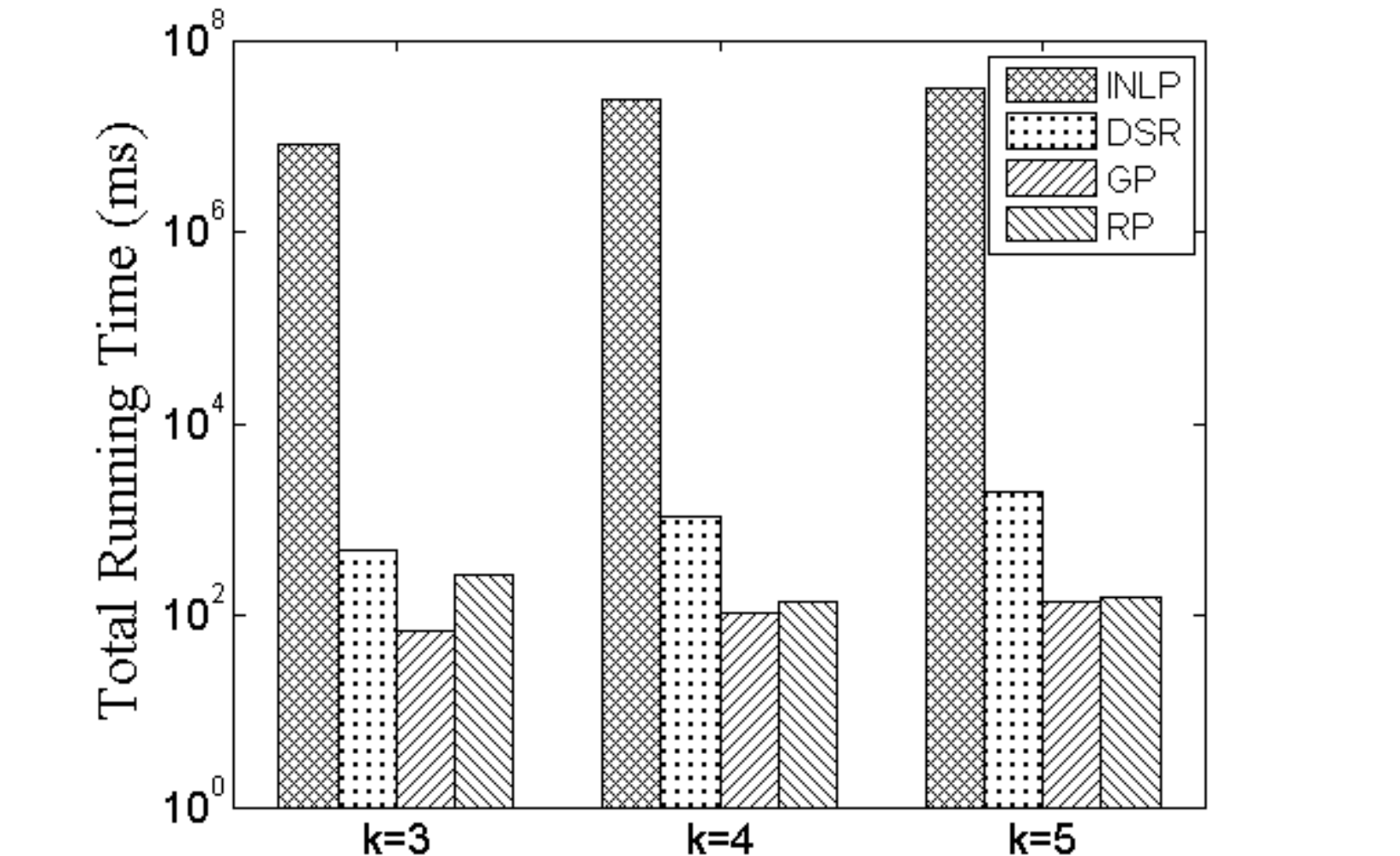}\label{Fig:TimeH3}}
\caption{Simulation results over 100 requests when $H=3$ for $16$-node network: (a) Acceptance Ratio (AR), (b) Average Number of Used Nodes (ANUN) and (c) Running Time.}%
\label{Fig:SimkH3}%
\end{figure*}

\begin{figure*}[tbh]
\centering
\subfloat[Acceptance Ratio (AR)]{
\includegraphics[trim=25mm 89mm 25mm 88mm,clip=true,width=0.33\textwidth]{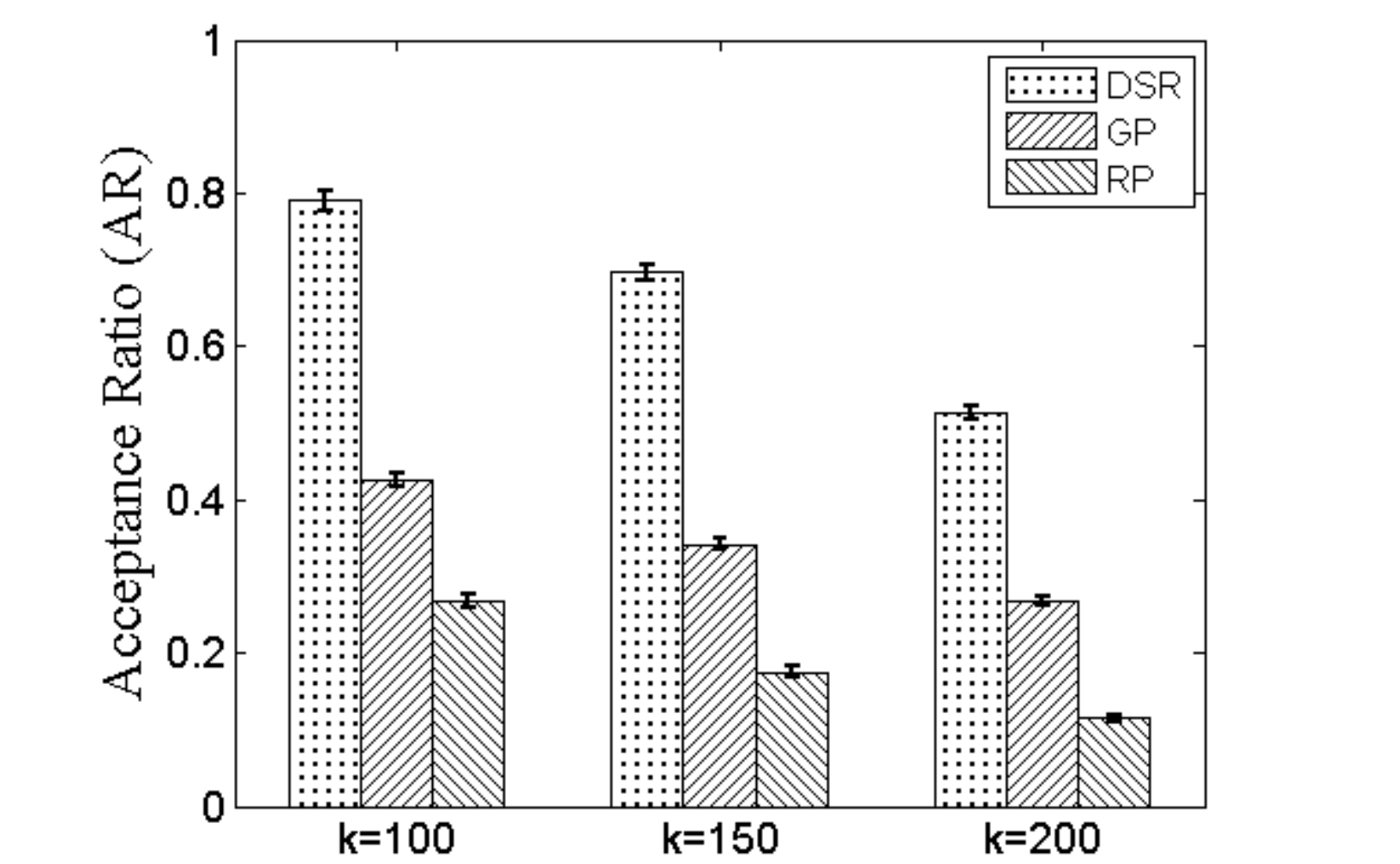}\label{Fig:AR100}}
\subfloat[Average Number of Used Nodes]{
\includegraphics[trim=25mm 89mm 25mm 88mm,clip=true,width=0.33\textwidth]{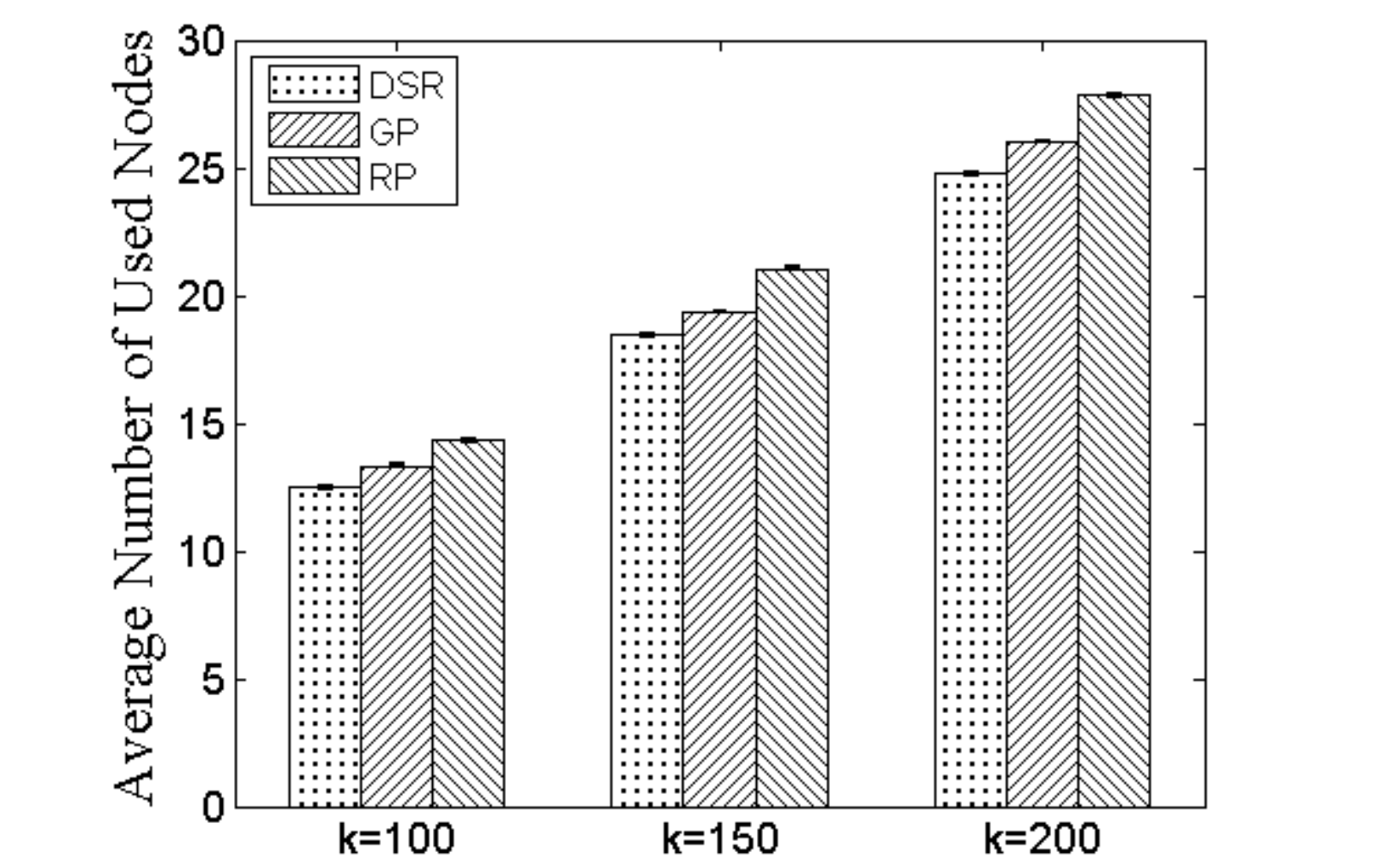}\label{Fig:ANUN100}}
\subfloat[Running Time]{
\includegraphics[trim=25mm 89mm 25mm 88mm,clip=true,width=0.33\textwidth]{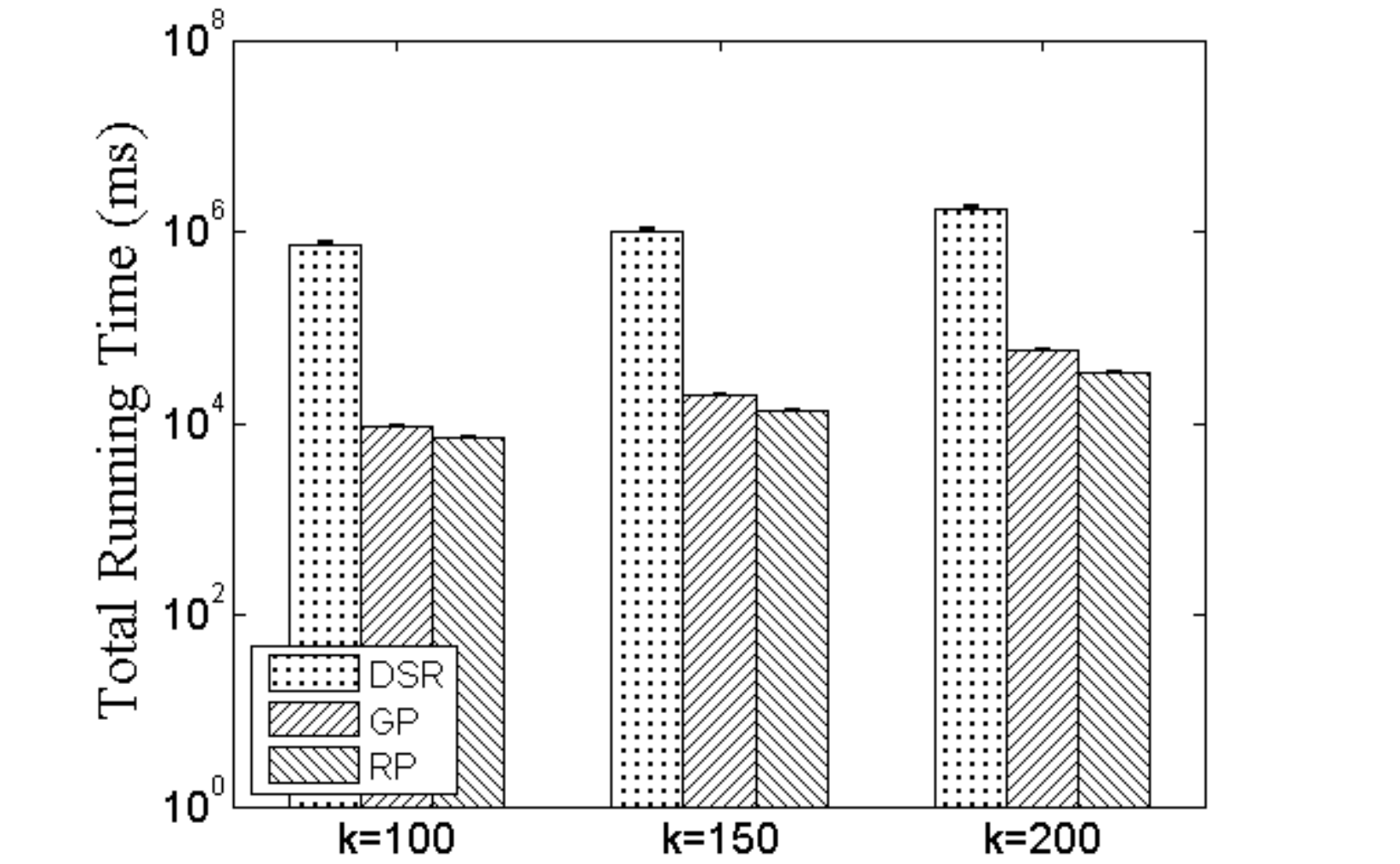}\label{Fig:Time100}}
\caption{Simulation results over 100 sets of 100 requests ($95\%$ confidence interval) when $H=2$ for $100$-node network: (a) Acceptance Ratio (AR), (b) Average Number of Used Nodes (ANUN) and (c) Running Time.}%
\label{Fig:SimLH2}%
\end{figure*}
\begin{figure*}[tbh]
\centering
\subfloat[Acceptance Ratio (AR)]{
\includegraphics[trim=25mm 90mm 25mm 89mm,clip=true,width=0.33\textwidth]{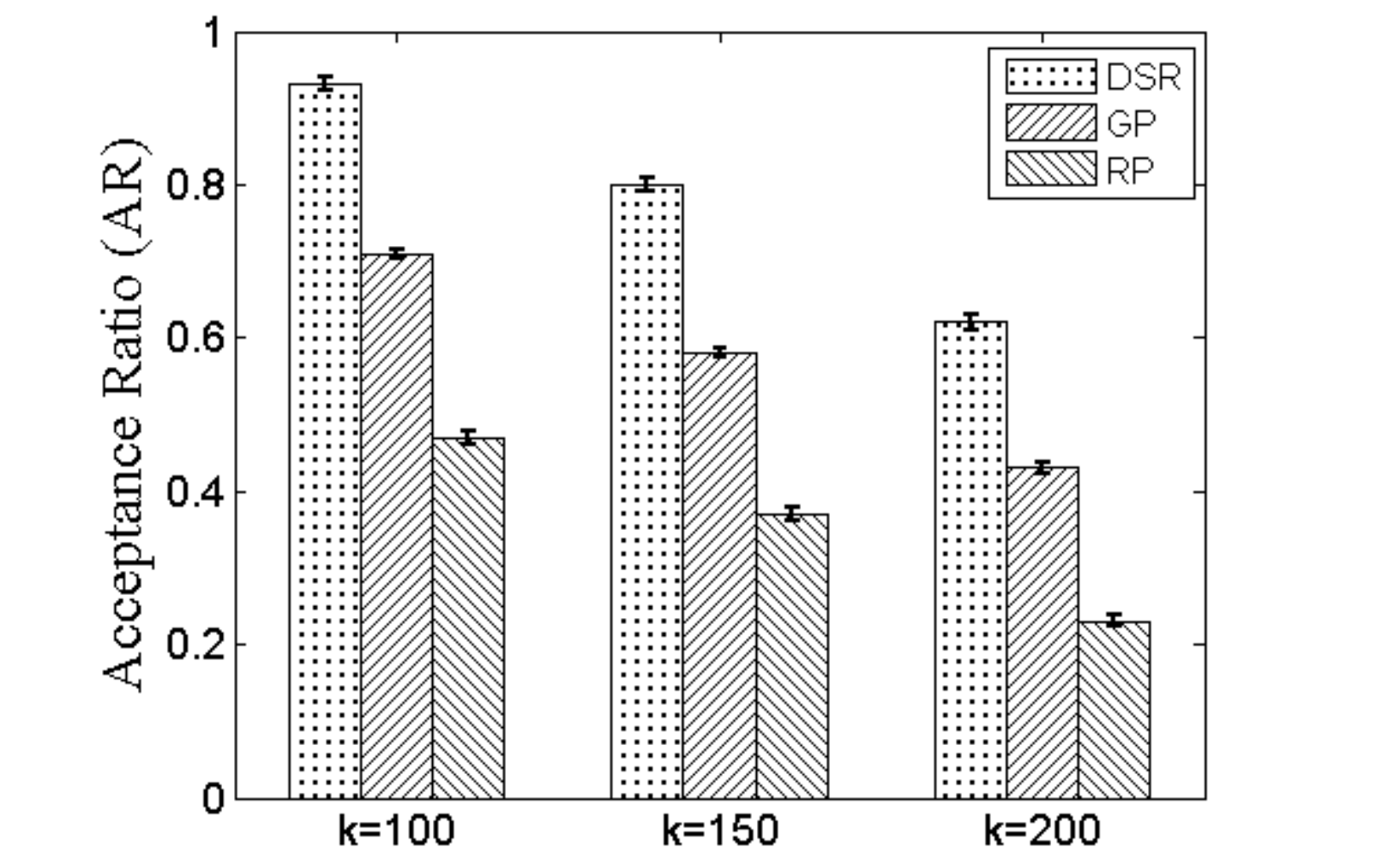}\label{Fig:AR150}}
\subfloat[Average Number of Used Nodes]{
\includegraphics[trim=25mm 90mm 25mm 89mm,clip=true,width=0.33\textwidth]{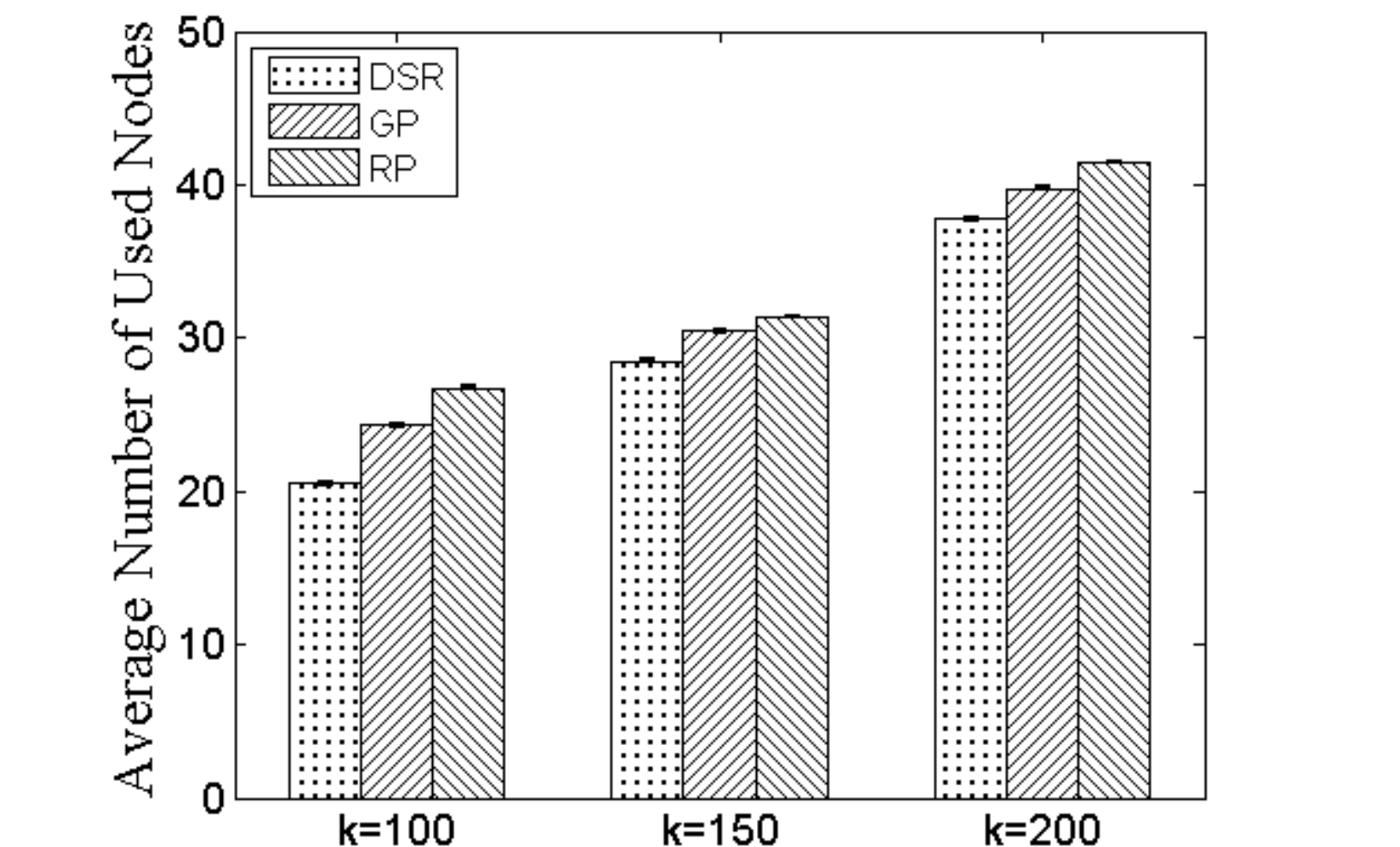}\label{Fig:ANUN150}}
\subfloat[Running Time]{
\includegraphics[trim=25mm 90mm 25mm 89mm,clip=true,width=0.33\textwidth]{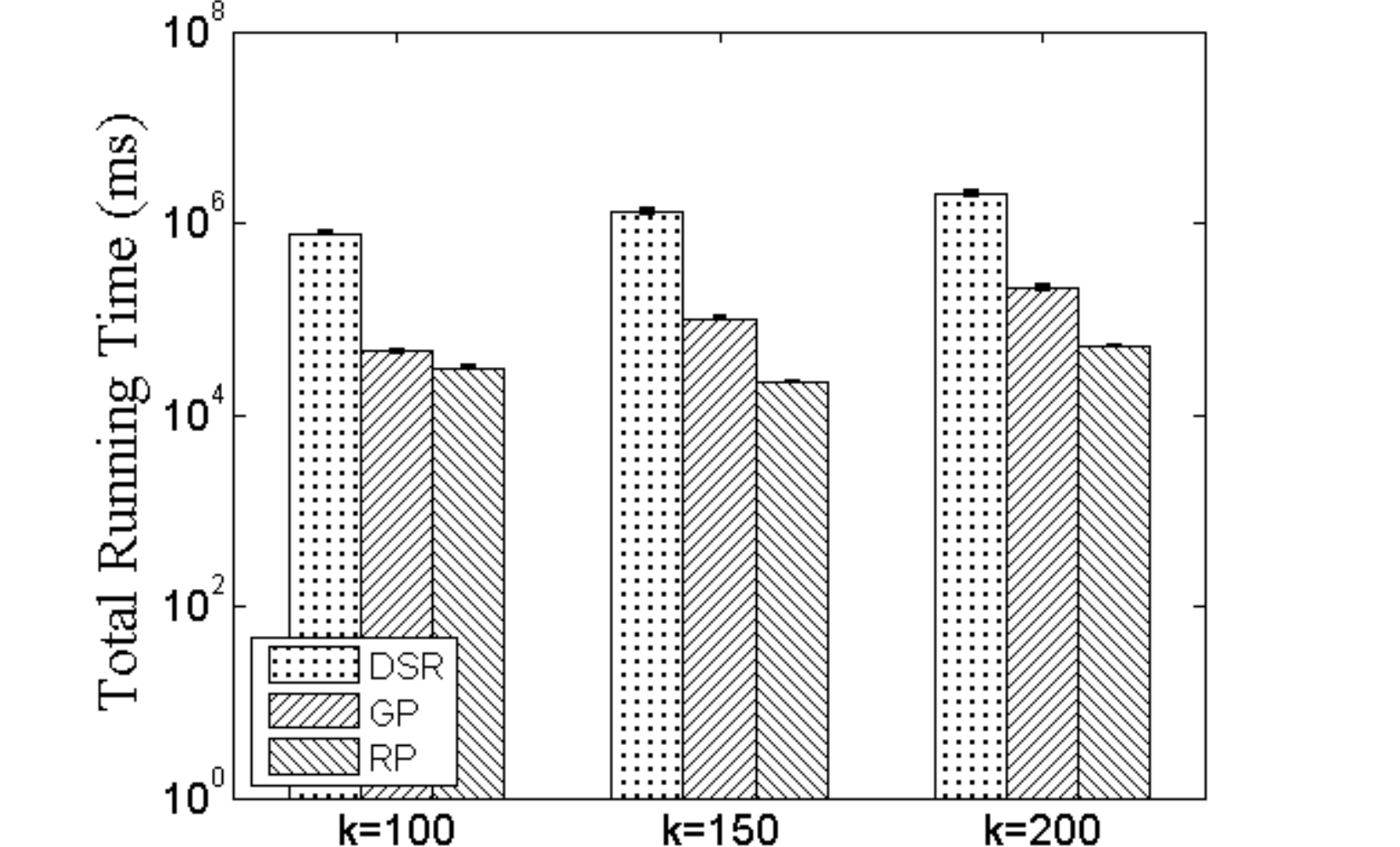}\label{Fig:Time150}}
\caption{Simulation results over 100 sets of 100 requests ($95\%$ confidence interval) when $H=3$ for $100$-node network: (a) Acceptance Ratio (AR), (b) Average Number of Used Nodes (ANUN) and (c) Running Time.}%
\label{Fig:SimLH3}%
\end{figure*}
\begin{figure*}[tbh]
\centering
\subfloat[Acceptance Ratio (AR)]{
\includegraphics[trim=25mm 90mm 25mm 89mm,clip=true,width=0.33\textwidth]{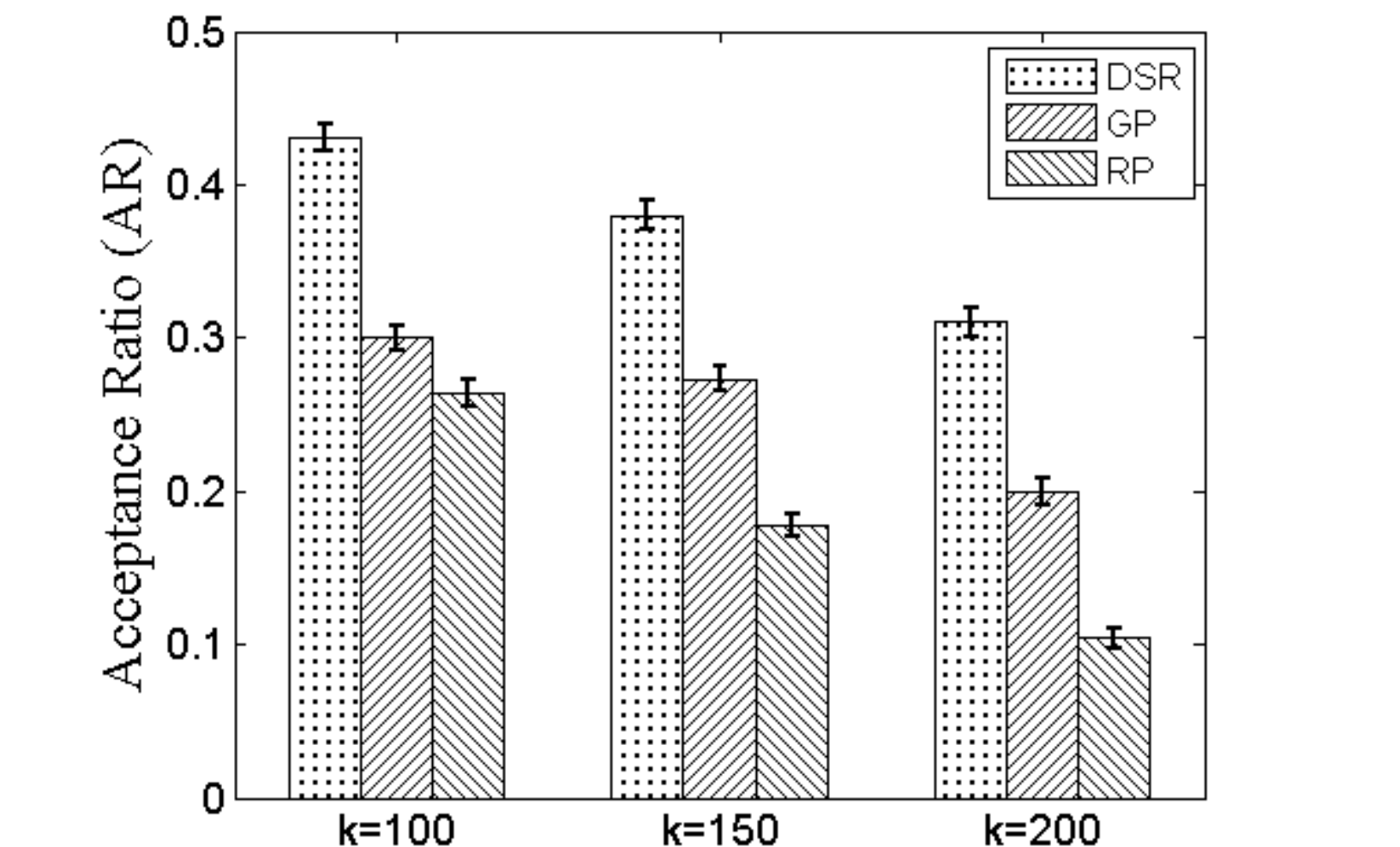}\label{Fig:ARSRNG}}
\subfloat[Average Number of Used Nodes]{
\includegraphics[trim=25mm 90mm 25mm 89mm,clip=true,width=0.33\textwidth]{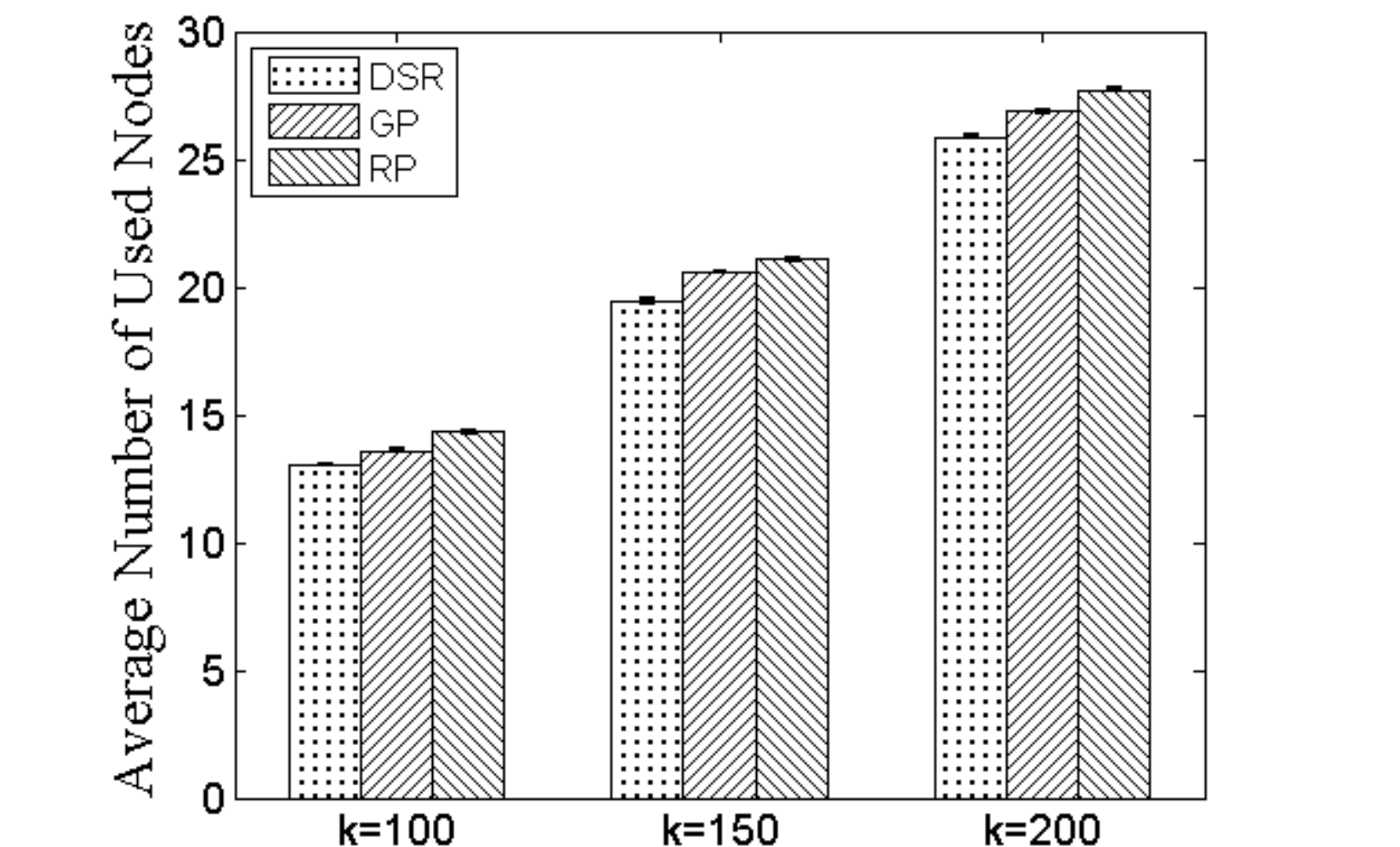}\label{Fig:ANUNSRNG}}
\subfloat[Running Time]{
\includegraphics[trim=25mm 90mm 25mm 89mm,clip=true,width=0.33\textwidth]{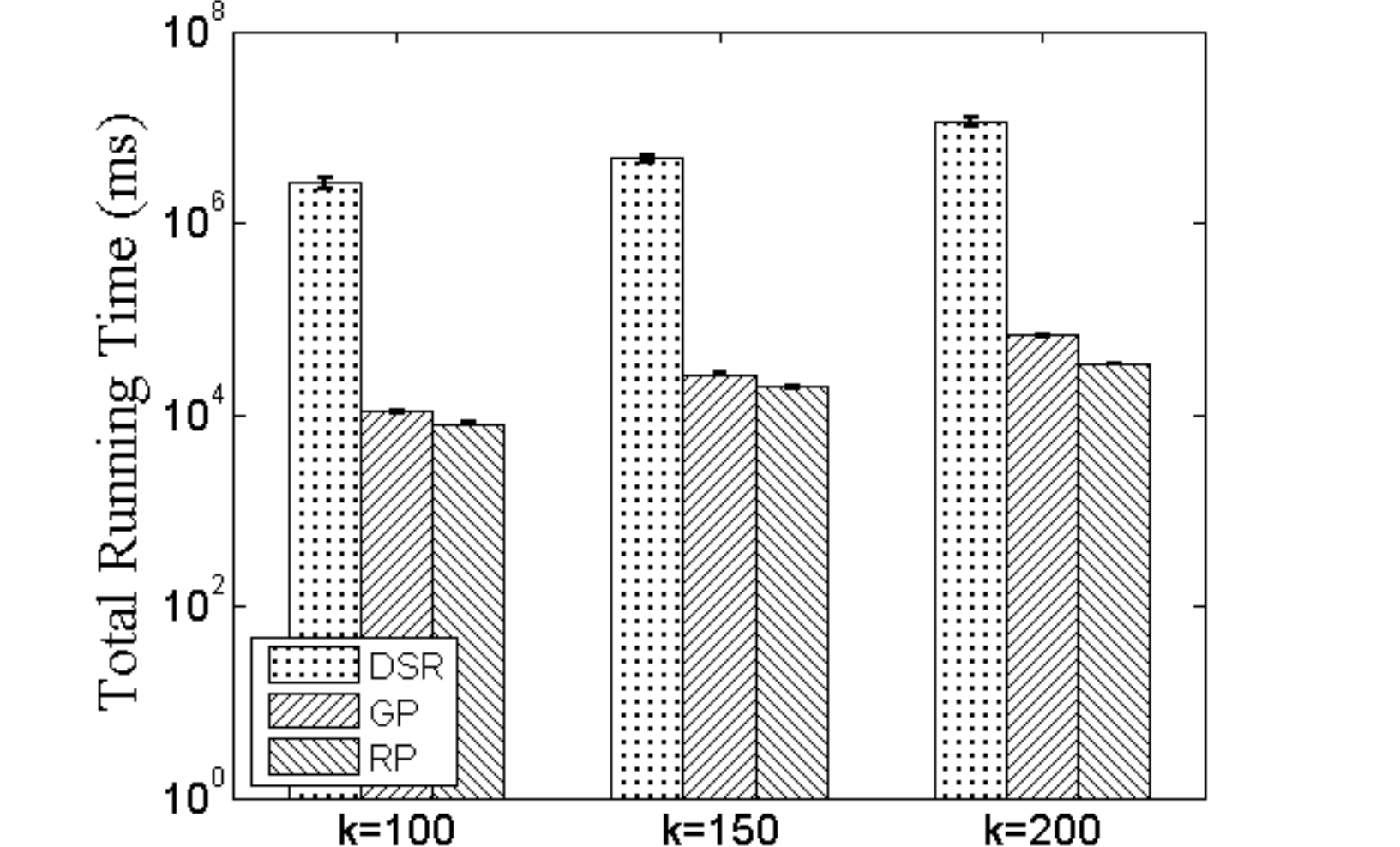}\label{Fig:TimeSRNG}}
\caption{Simulation results over 100 sets of 100 requests ($95\%$ confidence interval) when $H=2$ for $100$-node network with SRNG failures: (a) Acceptance Ratio (AR), (b) Average Number of Used Nodes (ANUN) and (c) Running Time.}%
\label{Fig:SimSRNG}%
\end{figure*}
We first conduct simulations on a 16-node network. If we set $c$ (VM demanding capacity) relatively too small, then by placing as many VMs as possible on one node it may return a solution. If we set $c$ relatively too big, then the solution may not exist. Therefore, we let the node's capacity be at most three times of the requested bandwidth of one single VM, by which we want to challenge the algorithm to find the solution. Consequently, the simulation parameters are set like this: the node capacities are randomly distributed between $100$ and $200$ units, and the node availabilities are randomly distributed among the set $\{0.99, 0.999, 0.9995, 0.9999 \}$. 
For each request $r(k,c,V, T,A,\delta)$, $k \in \left[3,5\right]$, $c \in \left[60, 130 \right]$, each element in the delay matrix $T$ is between $15$ and $25$, each element in the connection availability matrix $A$ is among the set $\{0.999, 0.9999\}$, and $\delta$ is in the set $\{0.999, 0.9999, 0.99999, 0.999999 \}$. We randomly generate $100$ requests for $k=3,4,5$, respectively. 
With respect to $F(m,n,\eta, D)$, when $\eta=0.999$ for link $(m,n)$, it returns $1$ when the delay is at most $D$, where $D$ is randomly chosen between $[10, 20]$, otherwise it returns $0$; when $\eta=0.9999$ for link $(m,n)$, it returns $1$ when the delay is at most $D$, where $D$ is randomly chosen between $[20, 30]$, otherwise it returns $0$.  
We set $H=2$ and $3$.

We first evaluate the performance of the algorithms in terms of Acceptance Ratio (AR), which is defined as the number of accepted requests over all the requests (between $0$ and $1$). Figs.~\ref{Fig:SimkH2}\subref{Fig:ARH2} and \ref{Fig:SimkH3}\subref{Fig:ARH3} show that the exact INLP always achieves the highest AR. DSR has a close performance with INLP, and it outperforms the other two heuristics. Besides, we notice that for the same algorithm, it achieves higher AR value when $H$ increases, since more VM replicas are allowed to be placed for a higher $H$. 


Next, we compare the algorithms in terms of Average Number of Used Nodes (ANUN). The ANUN is defined as the total number of nodes consumed by all the accepted requests divided by the number of accepted requests. From Figs.~\ref{Fig:SimkH2}\subref{Fig:ANUNH2} and \ref{Fig:SimkH3}\subref{Fig:ANUNH3}, we see that the achieved ANUN value by RP when $k=4, 5, H=2$, and when $k=5, H=3$ is the (or second) lowest. This is because its acceptance ratio in those scenarios is too low (under $15\%$), and it only finds solutions for some ``easier'' requests. Except for those cases, the INLP achieves the minimum value of ANUN, and our proposed DSR obtains the second lowest ANUN value. RP obtains a lower ANUN value than GP when $k=4$ and $H=3$, since it is regarded to place more shared VMs. From above, we observe that even under the constrained simulation setup, the exact INLP can always accept most requests and consume the least amount of nodes as well, which validates its correctness.

Finally, Figs.~\ref{Fig:SimkH2}\subref{Fig:TimeH2} and \ref{Fig:SimkH3}\subref{Fig:TimeH3} present the total running time over $100$ requests (in log scale). The INLP is significantly more time-consuming than all the 3 heuristics. The DSR, on the other hand, has a slightly higher running time than the other two heuristics, but it pays off by having a higher AR as shown in Figs.~\ref{Fig:SimkH2}\subref{Fig:ARH2} and \ref{Fig:SimkH3}\subref{Fig:ARH3}, and lower ANUN shown in Figs.~\ref{Fig:SimkH2}\subref{Fig:ANUNH2} and \ref{Fig:SimkH3}\subref{Fig:ANUNH3}. 

\subsubsection{$100$-node network without SRNG failures} \label{Sec:SimuLarge}

In this subsection, we simulate a $100$-node complete graph, where the node capacities are randomly distributed in $[1000, 2000]$. The other simulation setup follows the same with Section~\ref{Sec:SimuSmall}. Since the problem sizes increase largely, the INLP becomes very time-consuming and it keeps running for at least one day without returning a feasible solution. We therefore only evaluate the heuristic algorithms. Due to the lack of the INLP, we generate $100$ sets of $100$ traffic requests for each $k=100, 150, 200$ requested VMs, respectively, and evaluate all the heuristic algorithms for those $100$ sets of $100$ traffic requests ($100$ runs). By doing this, we want to establish confidence on the performance of heuristics. 
Figs.~\ref{Fig:SimLH2} and \ref{Fig:SimLH3} depict the AR, ANUN and running time (in log scale) of all these algorithms, where the confidence interval is set to $95\%$. The $95\%$ confidence interval is calculated for all the figures, but in those where it is not visible, the interval is negligibly small\footnote{We note here that some plots are log-scale that additionally contributes to the confidence interval visibility.}. 
Similar to Section~\ref{Sec:SimuSmall}, DSR always achieves better performance than the other 2 heuristics in terms of AR (see Figs.~\ref{Fig:SimLH2}\subref{Fig:AR100} and \ref{Fig:SimLH3}\subref{Fig:AR150}) and ANUN (see Figs.~\ref{Fig:SimLH2}\subref{Fig:ANUN100} and \ref{Fig:SimLH3}\subref{Fig:ANUN150}). On the other hand, Figs.~\ref{Fig:SimLH2}\subref{Fig:Time100} and \ref{Fig:SimLH3}\subref{Fig:Time150} show that DSR is more time consuming than the other heuristics, but it is still acceptable since it acquires higher AR and lower ANUN values. Another observation is that in this larger network scenario, RP obtains the highest ANUN value. This reveals that RP performs more poorly because of its randomness when the problem size grows. 

\subsubsection{$100$-node network with SRNG failures} \label{Sec:SimuSRNG}

It is assumed that there are in total $15$ SRNG events, and each SRNG event occurs with the probability in the set $\{0.000001,0.000002, 0.000003, 0.000004, 0.000005\}$. Each server node is associated with at most $5$ SRNG events. The other simulation setup follows the same with Section~\ref{Sec:SimuLarge} and we also evaluate all the three algorithms by 100 runs to establish confidence. Since more SRNG events are induced for each node, the total VM placement availability for the same set of nodes will decrease according to Eq.~(\ref{Eq:SRNG}), causing the optimal solution not to exist for when $\delta>0.9999$. 
Due to space limits, we only present the results for $H=2$ in Fig.~\ref{Fig:SimSRNG}, where a confidence interval is set to $95\%$. Similar to Sections~\ref{Sec:SimuSmall} and \ref{Sec:SimuLarge}, DSR can obtain a better performance than the other 2 heuristics in terms of AR (see Fig.~\ref{Fig:SimSRNG}\subref{Fig:ARSRNG}) and ANUN (see Fig.~\ref{Fig:SimSRNG}\subref{Fig:ANUNSRNG}), but this comes at the expense of a higher running time as shown in Fig.~\ref{Fig:SimSRNG}\subref{Fig:TimeSRNG} (in log scale). Due to the reason for incurring SRNG events, for all three algorithms, we can see that the achieved AR value in Fig.~\ref{Fig:SimSRNG}\subref{Fig:ARSRNG} is lower than Fig.~\ref{Fig:SimLH2}\subref{Fig:AR100}, and the obtained ANUN value in Fig.~\ref{Fig:SimSRNG}\subref{Fig:ANUNSRNG} is higher than Fig.~\ref{Fig:SimLH2}\subref{Fig:ANUN100}. 

In all, we conclude that the exact INLP can be used as an optimal solution when the computation speed is not a big concern. However, as the problem size increases, its running time will increase exponentially. On the contrary, our proposed DSR is a good compromise between performance and running time, and it is the preferred choice for when the VM placement request needs to be computed on-the-fly.

\section{Availability-Based Delay-Constrained Routing Problem} \label{Sec:Routing} 

In the RVMP problem, we do not consider the link delay and availability, and assume that the function $F(m,n,\eta, D)$ is given.   
In this section, we study how to find a connection over at most $w$ link-disjoint paths between a node pair, such that the connection availability is no less than $\eta$ and each path delay is no more than $D$. 
For completeness, let us first formulate the connection availability calculation, which is introduced in \cite{RNDM,Yang15}.

\subsection{Link Failure Scenarios}
Analogous to the node failure scenarios, we also address two kinds of link failure scenarios. For simplicity, suppose there are two (fully) link-disjoint paths $\psi_1$ and $psi_2$, and the availability of link $l$ is denoted as $A_l=1-f_l$, where $0<A_l \leq 1$ and $f_l$ is the failure probability of link $l$. Then their total availability $A^2_{FD}$ can be computed based on the following scenarios:

\begin{itemize}
\item \textbf{Single-link failure}: Analogous to the single node failure, here a path $\psi$'s availability (denoted by $A_{\psi}$) is equal to its lowest traversed link availability (highest failure probability). Using two disjoint paths will therefore lead to a total connection availability of $1$. In Appendix~B, we will address the ABDCR problem under the single-link failure scenario. 

\item \textbf{Multiple link failures:} This is a more general scenario where at one certain point in time, several links in the network may fail simultaneously. Hence, for a path $\psi$, its availability $A_{\psi}$ should take into account all its links' availabilities, i.e., $A_{\psi}=\prod_{l \in \psi } A_{l}$. Consequently, $A^2_{FD}=1-(1-A_{\psi1})(1-A_{\psi2})$, which indicates the probability that at least one of the two disjoint paths is available. In this paper, we assume multiple link failures may occur at any particular time point.   

\end{itemize}

\subsection{Connection Availability and Problem Definition}
Similar to the node availability, we assume that the link availability is equal to the product of availabilities of all its components (e.g., amplifiers). If a path $p$ contains the links $l_{1}$, $l_{2}$, $l_{3}$,\ldots, $l_{m}$, and their corresponding (independent) availabilities are denoted by $A_{l_1}$, $A_{l_2}$,
$A_{l_3}$,\ldots, $A_{l_m}$, then the availability of this (unprotected) path (represented by
$A_{\psi}$) is equal to $A_{\psi}=A_{l_1}\cdot A_{l_2}\cdot A_{l_3} \cdot \cdots \cdot A_{l_m}$. If we take the $-\log$ of the link availabilities, finding a path with the highest availability is equivalent to the shortest path problem \cite{Dijkstra59}. 

When, for a single connection (i.e., a single path), there are $w \geq 2$ paths $\psi_{1}$, $\psi_{2}$,\ldots, $\psi_{w}$ with availabilities represented by $A_{\psi_{1}}$, $A_{\psi_{2}}$,\ldots, $A_{\psi_{w}}$, the connection availability indicates the probability that at least one path is operational. We consider two cases, namely: (1) fully link-disjoint paths: these $w$ paths have no links in common, and (2) partially link-disjoint paths: at least two of these $w$ paths traverse a common link. In case (1), the availability (represented by $A^{w}_{FL}$) can be calculated as follows:

\begin{flalign}
A^{w}_{FL} =&1-\prod_{i=1}^{w} (1-A_{\psi_{i}}) 
\label{Eq. KFDisjointPath}
\end{flalign}

If we use Eq. (\ref{Eq. KFDisjointPath}) to calculate the availability for the partially link-disjoint case, the probability that the overlapping links operate (or the availability of the overlapping links) will be counted more than once. To solve this, we can analogously apply the operators $\circ$ and $\coprod$ introduced in Section~\ref{Sec:VMReliability}. Assuming there are $w$ partially link-disjoint paths $\psi_1$, $\psi_2$,\ldots, $\psi_w$, the availability (represented by $A^{w}_{PL}$) of $w$ partially link-disjoint paths can be calculated as:

\begin{align}
A^{w}_{PL}&=1-\coprod_{i=1}^{w} (1-A_{\psi_{i}}) \label{Eq. KPDisjointPath}
\end{align}    

\begin{figure}[tbh]
\centering
\includegraphics[trim = 0mm 0mm 0mm 0mm,clip,width=0.37\textwidth]{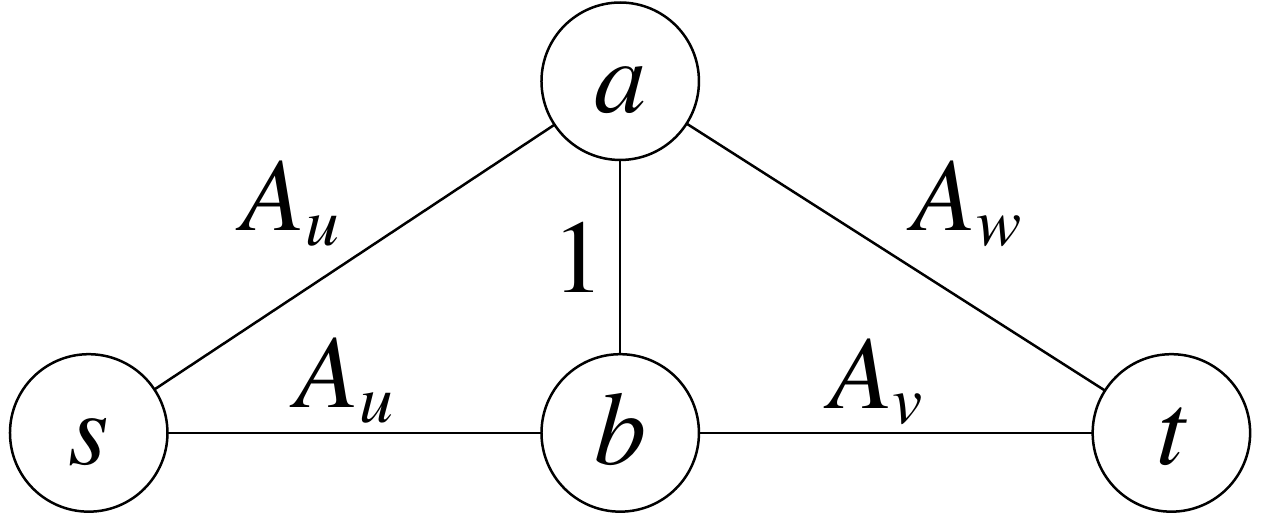}
\caption{Availability calculation of a pair of fully and partially link-disjoint paths.}%
\label{Fig:PDAvb}%
\end{figure}

Let us use an example to explain how to calculate the connection availability for fully and partially link-disjoint paths, where $w$ is set to $2$ for simplicity. In Fig. \ref{Fig:PDAvb} where the link availability is labeled on each link, paths $s-a-t$ and $s-b-t$ are fully link disjoint. According to Eq. (\ref{Eq. KFDisjointPath}), their availability is equal to: 
\begin{align}
& 1-(1-A_{u} \cdot A_{w})\cdot(1-A_{u} \cdot A_{v})\nonumber  \\
=& 1-(1-A_u \cdot A_v-A_u \cdot A_w+\underline{A_u} \cdot A_w \cdot \underline{A_u} \cdot A_v)  \nonumber \\
=& A_{u} \cdot A_{v}+A_{u} \cdot A_{w}- \underline{A^{2}_{u}} \cdot A_{w} \cdot A_{v}
\end{align}

On the other hand, paths $s-a-t$ and $s-a-b-t$ are two partially link-disjoint paths. According to Eq. (\ref{Eq. KPDisjointPath}), the connection availability can be calculated as follows:

\begin{align}
& 1-(1-A_{u} \circ A_{w})\circ (1-A_{u} \circ A_{v})\nonumber  \\
=& 1-(1-A_u \circ A_v-A_u \circ A_w+\underline{A_u} \circ A_w \circ \underline{A_u} \circ A_v) \nonumber \\
= & A_{u} \cdot A_{v}+A_{u} \cdot A_{w}- \underline{A_{u}} \cdot A_{w} \cdot A_{v}
\end{align}

Next, we formally define the Availability-Based Delay-Constrained Routing (ABDCR) problem as follows:

\begin{definition}
Given is a network represented by $G(\mathcal{N},\mathcal{L})$, where $\mathcal{N}$ represents the set of $N$ nodes and $\mathcal{L}$ denotes the set of $L$ links. Each link $l \in \mathcal{L} $ is associated with an availability value $A_{l}$ and a delay value $d_l$. For a communication request represented by $r(s, t, \eta, D)$, where $s$ and $t$ denote the source and destination, $\eta$ $(0<\eta \leq 1)$ represents the connection availability requirement and $D$ indicates the delay constraint, the Availability-Based Delay-Constrained Routing (ABDCR) problem is to establish a connection over at most $w$ (partially) link-disjoint paths, such that the connection availability is at least $\eta$ and each path has a delay no more than $D$.
\end{definition}

In the ABDCR problem, we regard that each request corresponds to the communication between each VM pair which is resident on different nodes. 
When, the delay constraint is not imposed on each path, the ABDCR problem is equivalent to the Availability-Based Path Selection (ABPS) problem \cite{RNDM, Yang15}. In \cite{RNDM, Yang15} we have proved that the ABPS problem is NP-hard for $w \geq 2$ and cannot be approximated to an arbitrary degree. Therefore, the ABDCR problem for $w \geq 2$ is also NP-hard and cannot be approximated to an arbitrary degree. When $w=1$, the ABDCR problem is equivalent to the Multi-Constrained Routing problem, which is also NP-hard \cite{SAMCRA}. In the following, we propose an exact algorithm and two heuristics to solve the ABDCR problem.

\subsection{Exact Algorithm} \label{Sec:RoutingExactAlg}

To solve the ABDCR problem exactly, we apply a modified Dijkstra's algorithm by letting each node store as many subpaths as possible, which is similar to the exact algorithm for solving the multi-constrained routing problem \cite{SAMCRA}. We start with some notations used in the algorithm: \\

$sus[u][m]$: the parent node of node $u$ for its stored $m$-th subpath from $s$ to $u$. 

$avb[u][m]$: the availability value stored at node $u$ for its stored $m$-th subpath from $s$ to $u$. 

$delay[u][m]$: the delay value stored at node $u$ for its stored $m$-th subpath from $s$ to $u$.

$counter[u]$: the number of stored subpaths of node $u$. 

$sp[u][m]$: node $u$'s stored $m$-th subpath from $s$ to $u$. 

$adj(u)$: the set of adjacent nodes of node $u$.   \\

The pseudo code of the exact algorithm for solving the ABDCR problem when $w=1$ is given in Algorithm \ref{Alg: ExactABDCRW1}. 

\begin{algorithm}[h]
\caption{ABDCRw1$(G, s, t,\eta,D)$} \label{Alg: ExactABDCRW1}
$Q \leftarrow s$, $avb[s][1] \leftarrow 1$, $delay[s][1] \leftarrow 0$, $avb[i][m] \leftarrow +\infty$, $sus[i][m] \leftarrow i$, $counter[s] \leftarrow 1$, $counter[i] \leftarrow 0$, $ \forall i \in \mathcal{N} \backslash \{s\} $ \\
 \While{$Q \neq \emptyset $}
 {
  $u[m] \leftarrow$ Extract-min($Q$)\\
  \eIf{$u==t~ \&\& ~avb[u][m] \geq \eta$}
  {
   Return the path $sp[u][m]$\\
   
   }
   {
    \ForEach{$v \in adj(u)$}
    {
    	\If{$delay[u][m]+d_{uv} \leq D$}
    	{
    			$counter[v]=counter[v]+1$ \\
    			Assign the availability of subpath $sp[u][m]$-$(u,v)$ to $avb[v][counter(v)]$ \;
    			\small{$delay[v][counter(v)] \leftarrow delay[u][m]+d_{uv}$} \\
    			\normalsize{$sus[v][counter(v)] \leftarrow u$} \\
    			\normalsize{Insert ($Q,v,counter(v)$)} \\
    	}
    }
  }
 }

\end{algorithm}

\begin{figure}[tbh]
\centering
\includegraphics[trim = 0mm 0mm 0mm 0mm,clip,width=0.49\textwidth]{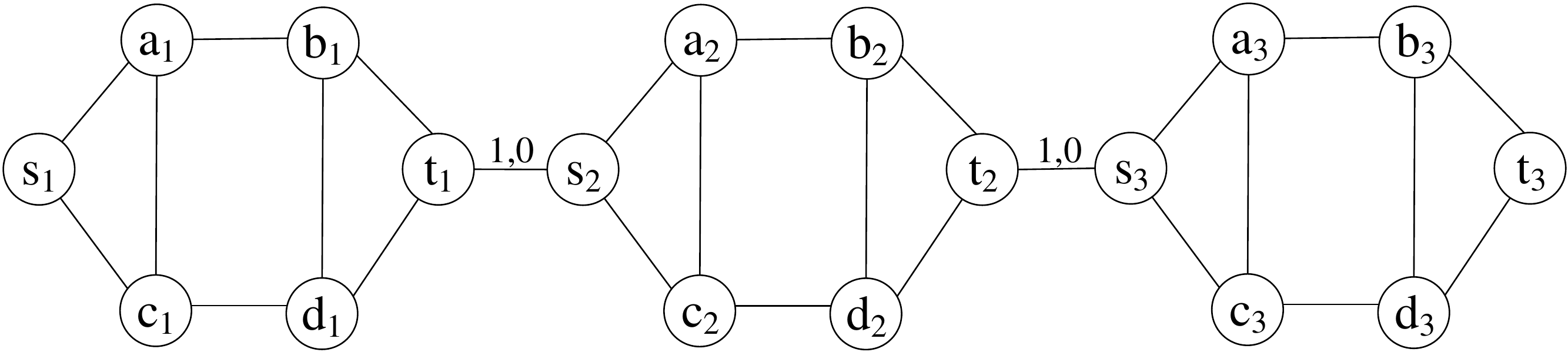}
\caption{An example of graph transformation for solving the ABDCR problem when $H=3$.}%
\label{Fig:ExactABDCRTransform}%
\end{figure}

When $w>1$ in the ABDCR problem, we could first duplicate the originate graph $\mathcal{G}$ into $w$ copies $\mathcal{G}^1$, $\mathcal{G}^2$,..., $\mathcal{G}^w$. After that, we create link $(t_i,s_j)$ to connect each graph copy with availability $1$ and delay $0$ for $1 \leq i \leq w$, $j=i+1$ except for $i=w$, where $n \in \mathcal{G}$ and $n_i \in \mathcal{G}^i$. By doing this, we obtain an auxiliary graph $\mathcal{G}^l$, where the source node is $s_1$ and the destination node(s) is $t_i$. In this context, if we find a path $\psi$ from $s_1$ to $t_i$ in $\mathcal{G}^l$, we can map $\psi$ into the original graph $\mathcal{G}$, where we can get $i$ (link-disjoint) paths from $s$ to $t$. For example, Fig.~\ref{Fig:ExactABDCRTransform} depicts the resulted auxiliary graph after transformation from the topology shown in Fig.~\ref{Fig:DPAvb} when $H=3$. Suppose we find a path $\psi_1$: $s_1$-$a_1$-$b_1$-$t_1$-$s_2$-$c_2$-$d_2$-$t_2$, then it is equivalent to say that we find two link-disjoint paths s-a-b-t and s-c-d-t. On the other hand, if there is a path $\psi_2$: $s_1$-$a_1$-$b_1$-$t_1$-$s_2$-$a_2$-$b_2$-$t_2$, then it can be mapped to the same path $s$-$a$-$b$-$t$ in the original graph. In this context, we regard that the availability of $\psi_2$ in $\mathcal{G}^l$ is the same with $s$-$a$-$b$-$t$ in $\mathcal{G}$.

As a result, we could slightly modify Algorithm~\ref{Alg: ExactABDCRW1} to solve the ABDCR problem for $w>1$ as follows ($1 \leq i \leq w$): 
\begin{itemize}

\item In Step $4$, the condition should only be \textbf{if} $u==t_i~\&\&~$the availability of $i$ link-disjoint paths (after mapping them to the original graph) is greater than or equal to $\eta$.

\item In Step $10$, for each node $v_i$, its $avb$ value is calculated based on $i$ link-disjoint paths after mapped to $\mathcal{G}$ together with the subpath from $s_i$ to $v_i$, according to Eq.~(\ref{Eq:KPDisjointAva}). For example, suppose in Step 3 node $a_2$ is extracted and its stored subpath is $s_1$-$a_1$-$b_1$-$t_1$-$s_2$-$a_2$. Now in Step 7, suppose the neighbor node $c_2$ is selected to update its $avb$ value according to node $a_2$. Therefore the subpath from $s_1$ to $c_2$ is $s_1$-$a_1$-$b_1$-$t_1$-$s_2$-$a_2$-$c_2$, which are paths s-a-b-t and s-a-c after mapped to the original graph. 

\item In step $11$, $s_i$ does not update its $delay$ value. It also sets $delay[s_i][counter(s_i)]=0$. 

\end{itemize}

It is worthwhile to mention that our proposed exact algorithm can also solve the ABDCR problem in Shared-Risk Link Group (SRLG) networks. Similar to SRNG, the links in the same SRLG will fail simultaneously if the group they belong to fails. For example, in optical networks \cite{Mukherjee06}, several fibers may reside in the same duct and a cut of the duct would cut all fibers in it. One duct in this context corresponds to one distinct SRLG.
To solve it, we only need to change the connection availability calculation in Step $4$ of Algorithm \ref{Alg: ExactABDCRW1}. More details of the connection availability in SRLG networks can be found in \cite{RNDM, Yang15}. 

The time complexity of algorithm~\ref{Alg: ExactABDCRW1} can be computed as follows. Let $Z_{\max}$ denote the maximum number of subpaths for each node to store, then in Step $2$, $Q$ contains at most $Z_{\max}N$ subpaths. According to \cite{VanMieghem01}, $Z_{\max} \leq \lfloor  e(N-2)! \rfloor$, where $e \approx 2.718$ is the Euler's number. When using a Fibonacci heap to structure the heap, selecting the minimum cost path has a time complexity of $O(\log(Z_{\max}N))$ \cite{Cormen01} in Step $3$. Step $7$-Step $13$ take at most $O(Z_{\max})$ time for each link to be iterated thus resulting in $O(Z_{\max}L)$ time; because for a fixed link, the steps within the inner loop (Steps $8$-$13$) all cost $O(1)$ time. Hence, the overall time complexity of Algorithm~\ref{Alg: ExactABDCRW1} is $O(Z_{\max}N \log(Z_{\max}N)+Z_{\max}L)$. Similarly, when $w>1$ for the exact algorithm, the overall time complexity is $O(Z_{\max}wN \log(Z_{\max}wN)+Z_{\max}wL)$.

\subsection{Heuristic algorithms} \label{Sec:RoutingHeu}
We propose two heuristic algorithms to solve the NP-hard ABDCR problem. The first one is called SeqTAMCRA: it leverages on TAMCRA \cite{SAMCRA}, which is a heuristic to solve the multi-constrained routing problem. The procedure of SeqTAMCRA is the following: it iteratively runs TAMCRA, so in each iteration, we may obtain a path with the biggest availability and delay no more than $D$. After each iteration, the traversed links will be pruned. This procedure continues until the connection availability is satisfied or the number of paths is bigger than $w$.

The second heuristic is called TADRA, Tunable Availability-based Delay-constrained Routing Algorithm, and it is identical to the exact algorithm except that the number (variable $counter$) of stored paths for each node should not exceed a given value (say $M$). For instance, Step $8$ of Algorithm~\ref{Alg: ExactABDCRW1} should be rewritten as: \newline
\textbf{if} $  counter[v]\leq M ~  \& \& ~ delay[u][m]+d_{uv} \leq D $ \textbf{then}. 

\subsection{Simulations}
\begin{figure}[tbh]
\centering\includegraphics[trim = 0mm 0mm 0mm 0mm,clip,width=0.25\textwidth]{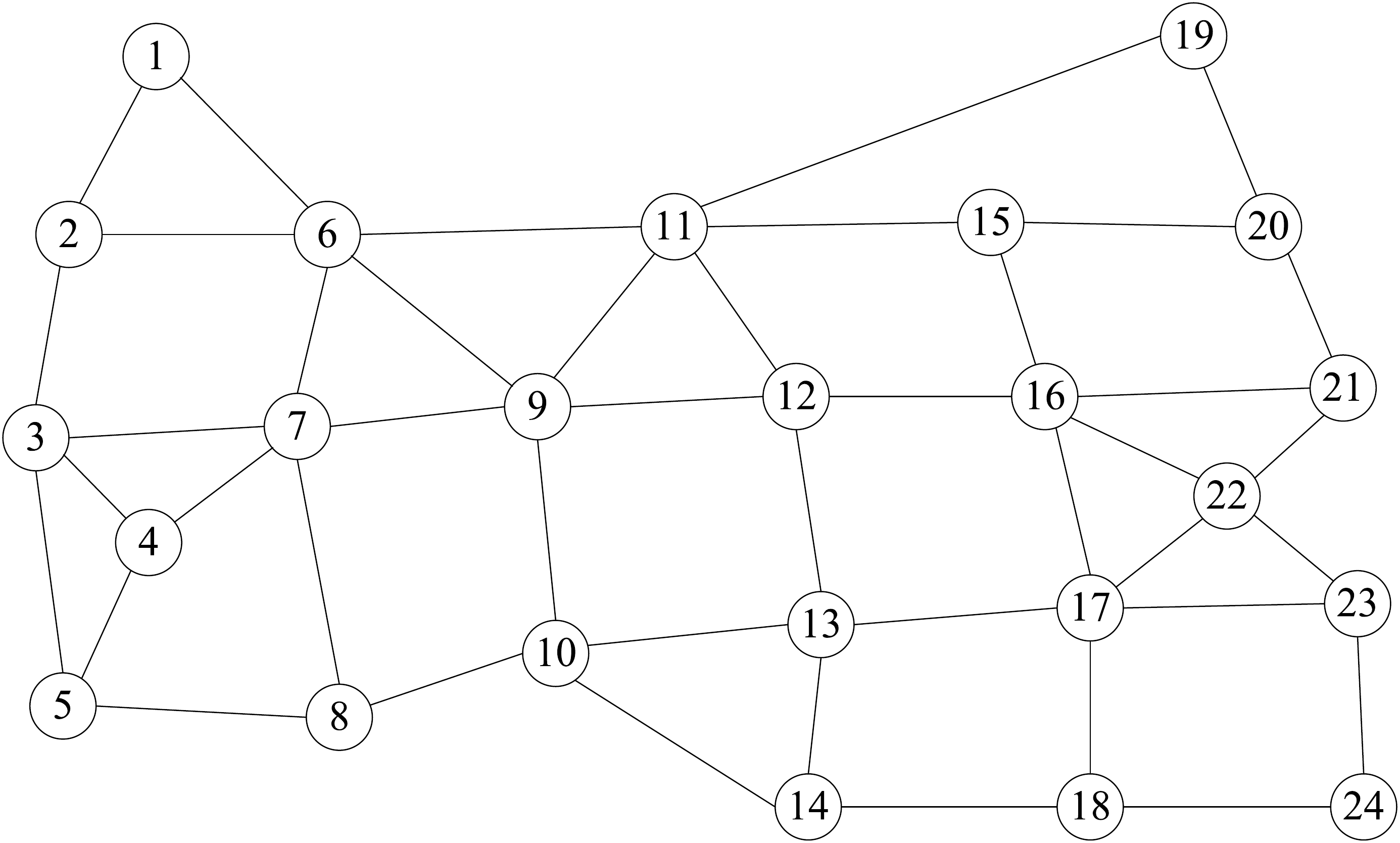}
\caption{USA carrier backbone network.}%
\label{Fig:USA}%
\end{figure}

\begin{figure}[tbh]
\centering\includegraphics[trim = 0mm 0mm 0mm 0mm,clip,width=0.3\textwidth]{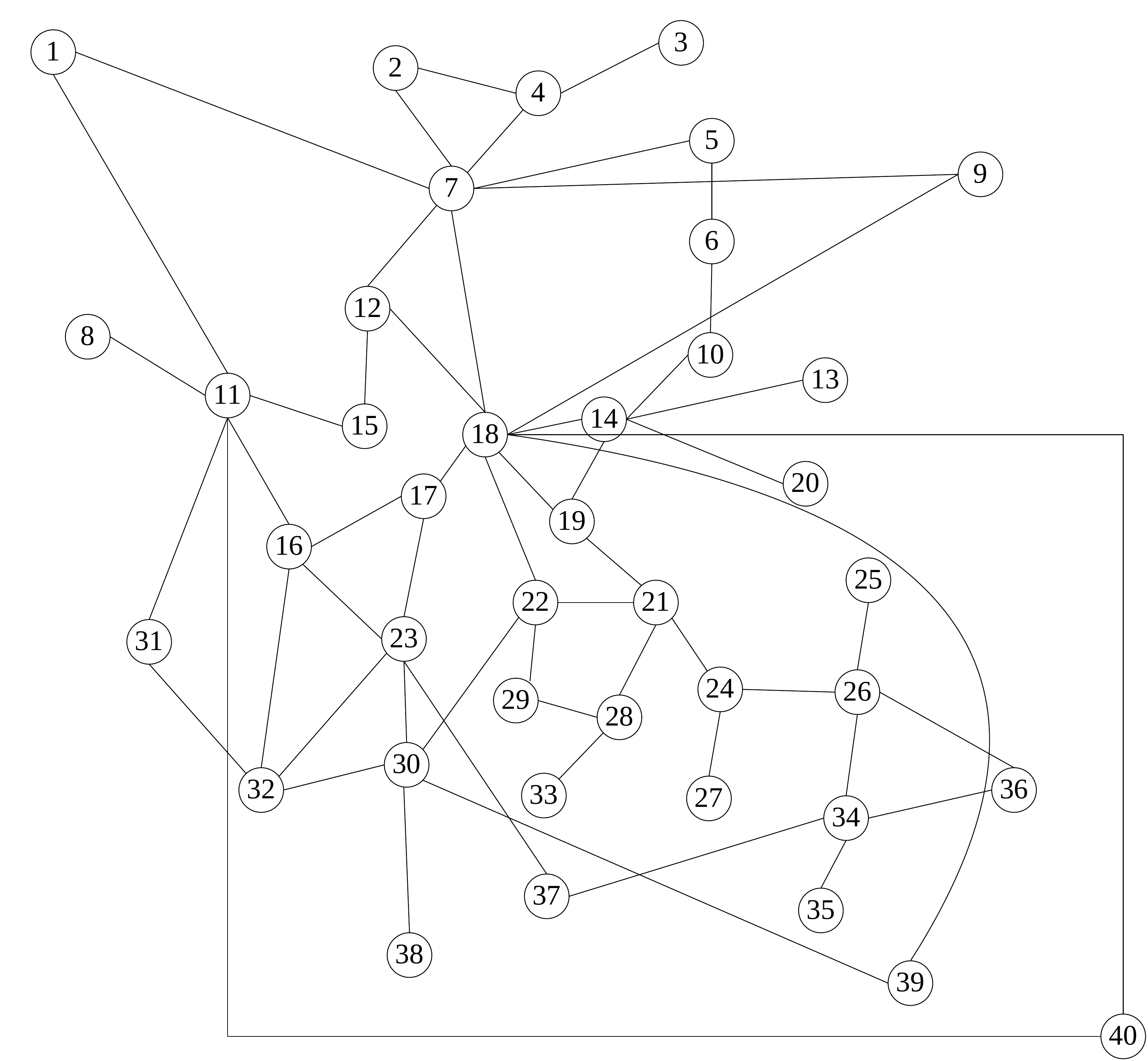}
\caption{G\'{E}ANT pan-European research network.}%
\label{Fig:GEANT}%
\end{figure}

In order to verify the proposed algorithms, we conduct simulations on two networks\footnote{Since most of typical data center network topologies are tree-like (e.g., Fat-Tree, BCube), the number of link-disjoint paths between node pairs is limited. Hence, to examine the algorithms more thoroughly, we choose well-connected backbone networks for the evaluation.}: USANet, displayed in Fig. \ref{Fig:USA}, which is a realistic carrier backbone network, and G\'{E}ANT, shown in Fig. \ref{Fig:GEANT}, which is a pan-European communications infrastructure. The link availability values are from the set $\{0.99, 0.999, 0.9999 \}$, and the link delays are set between $10$ and $25$. We randomly generate $1000$ requests, and for each request $r(s, t, \eta, D)$, $s$ and $t$ are randomly generated, $\eta$ is among the set $\{0.9995,0.9996,0.9997,0.9998,0.9999 \}$, and $D \in \left[15, 25 \right]$. We set $w=1,2,3$. For both SeqTAMCRA and TADRA, the maximum number of stored paths is set to $wN$, where $w$ is the number of maximum link-disjoint paths and $N$ is the number of nodes in the network. 

\begin{figure}[tbh]
\centering
\subfloat[USANet]{
\includegraphics[trim=25mm 90mm 25mm 90mm,clip=true,width=0.24\textwidth]{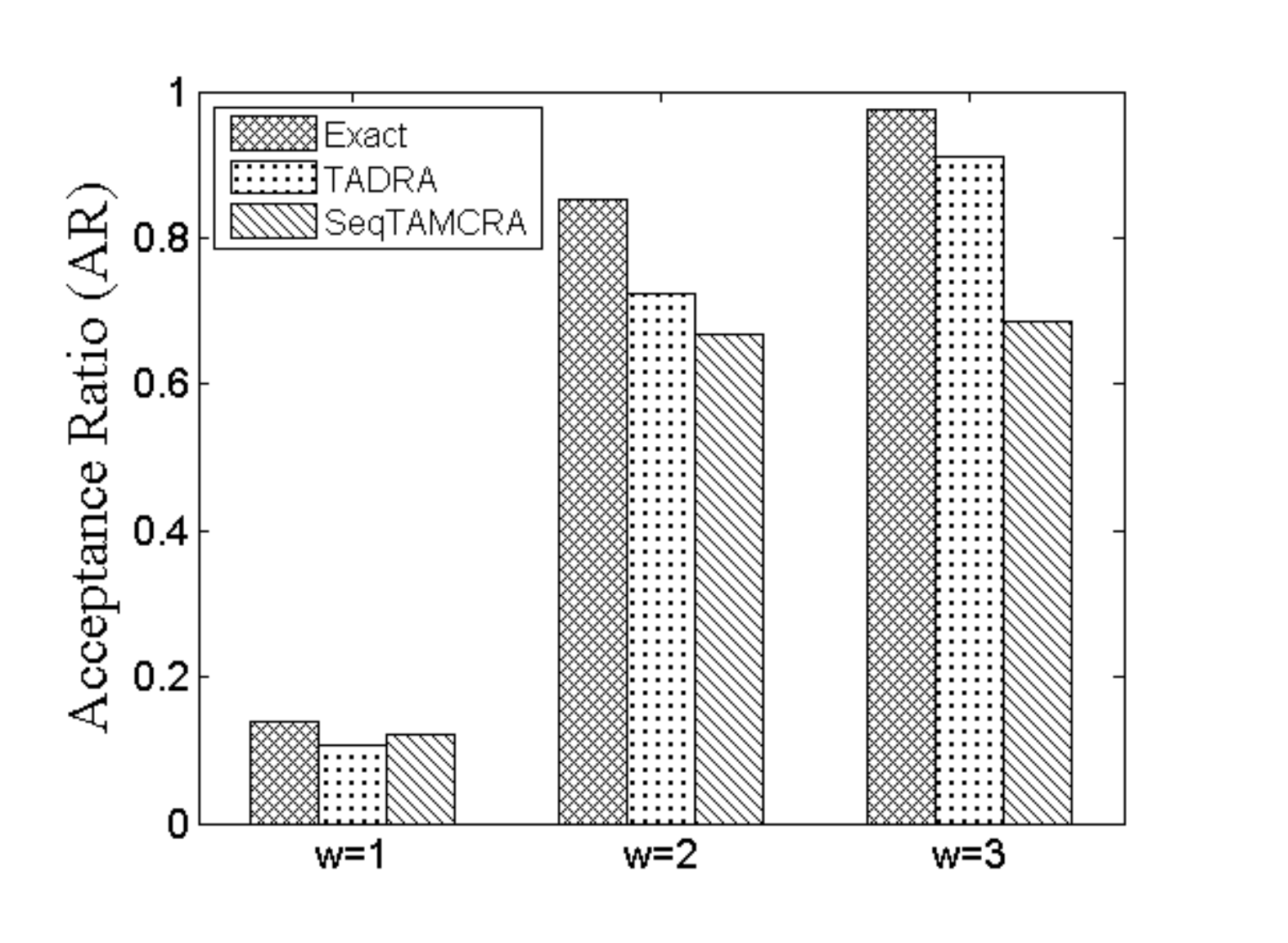}\label{Fig:ARUSA}}
\subfloat[G\'{E}ANT]{
\includegraphics[trim=25mm 90mm 25mm 90mm,clip=true,width=0.24\textwidth]{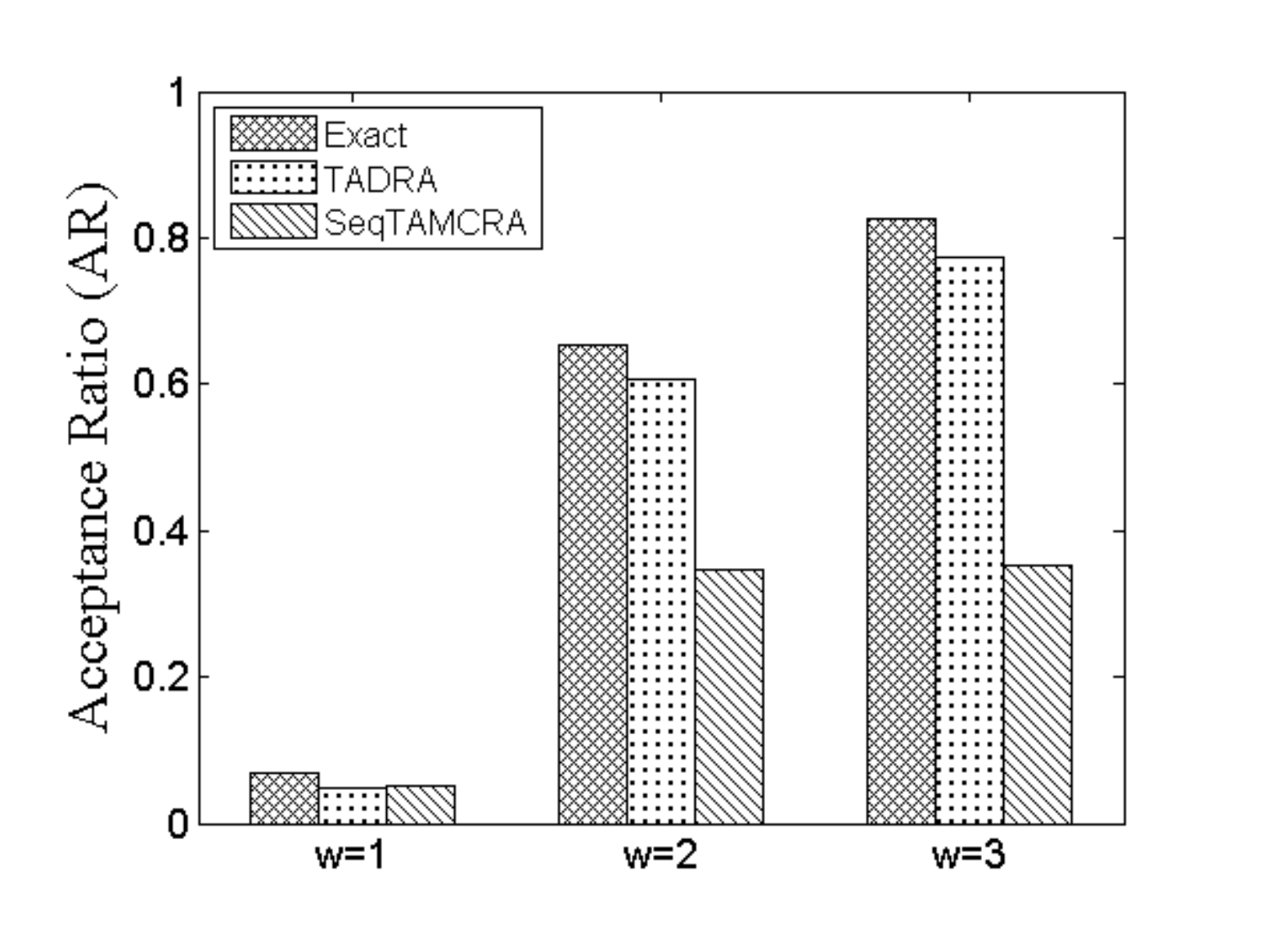}\label{Fig:ARGEANT}}
\caption{Acceptance ratio in two networks: (a) USANet (b) G\'{E}ANT .}%
\label{Fig:ABDCR_AR}%
\end{figure}

Figs.~\ref{Fig:ABDCR_AR}\subref{Fig:ARUSA} and \ref{Fig:ABDCR_AR}\subref{Fig:ARGEANT} depict the Acceptance Ratio (AR) of the 3 algorithms, which is defined by the number of accepted requests divided by the total number of requests. The exact algorithm can always achieve the highest AR value, which also verifies its correctness. TADRA obtains close to optimal performance when $w=2$ and $w=3$, while SeqTAMCRA performs well when $w=1$. The reason is that when $w=1$ SeqTAMCRA dynamically maintains $wN$ best paths for each node compared to the TADRA, so it can achieve a better performance. When $w>1$, after finding a feasible path by SeqTAMCRA, pruning the used links will prevent it to find link-disjoint paths in some cases, leading to worse performance. Moreover, dynamically maintaining fixed number of best paths only works for when $w=1$, since the connection availability calculation is non-linear when $w \geq 2$. Therefore, the exact algorithm and TADRA cannot adopt this technique to improve the efficiency for finding feasible paths. Nevertheless, we could jointly use TADRA when $w>1$ and SeqTAMCRA when $w=1$ as a heuristic combination.

\begin{figure}[tbh]
\centering
\subfloat[USANet]{
\includegraphics[trim=25mm 90mm 25mm 90mm,clip=true,width=0.24\textwidth]{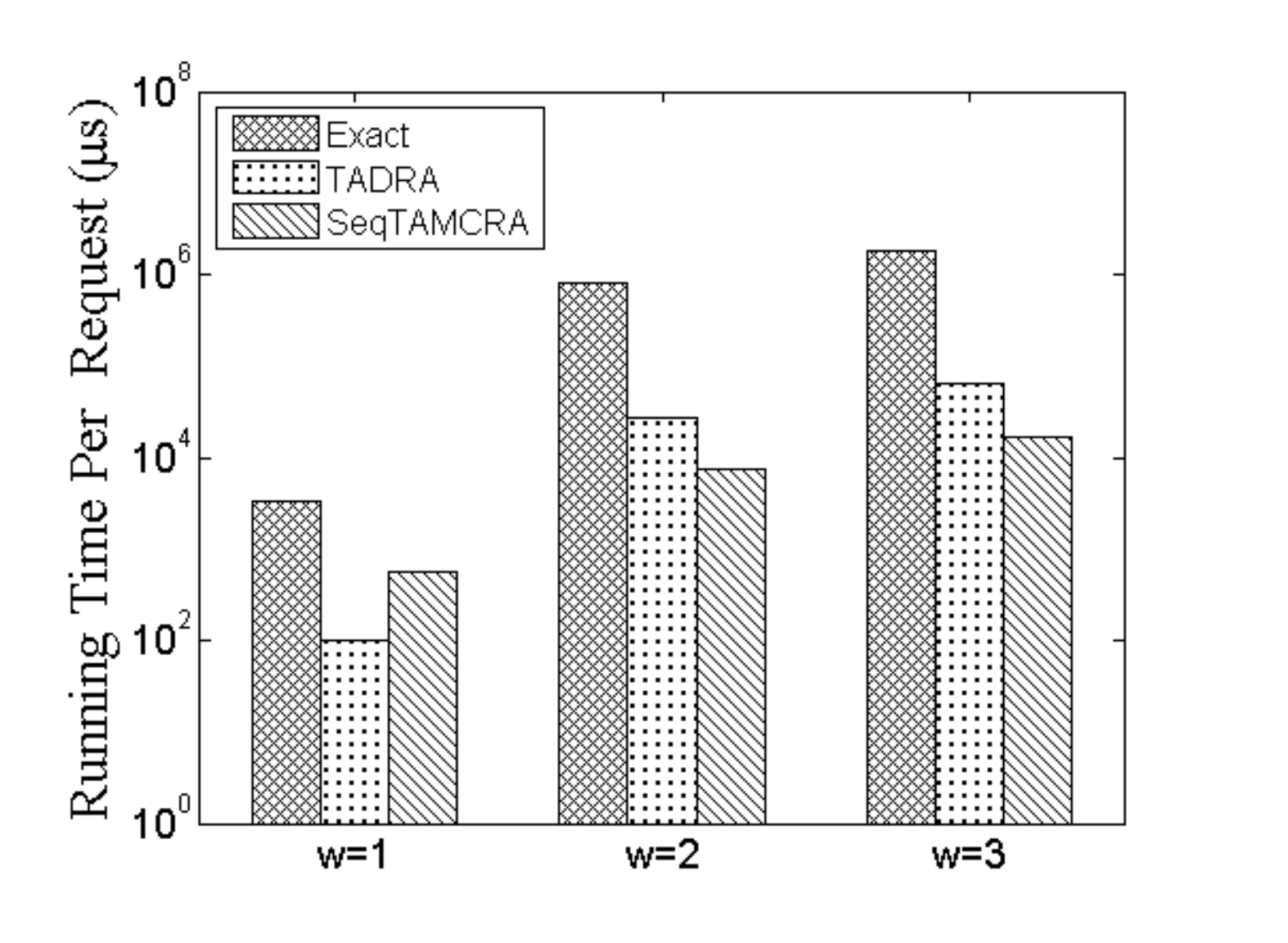}\label{Fig:TimeUSA}}
\subfloat[G\'{E}ANT]{
\includegraphics[trim=25mm 90mm 25mm 90mm,clip=true,width=0.24\textwidth]{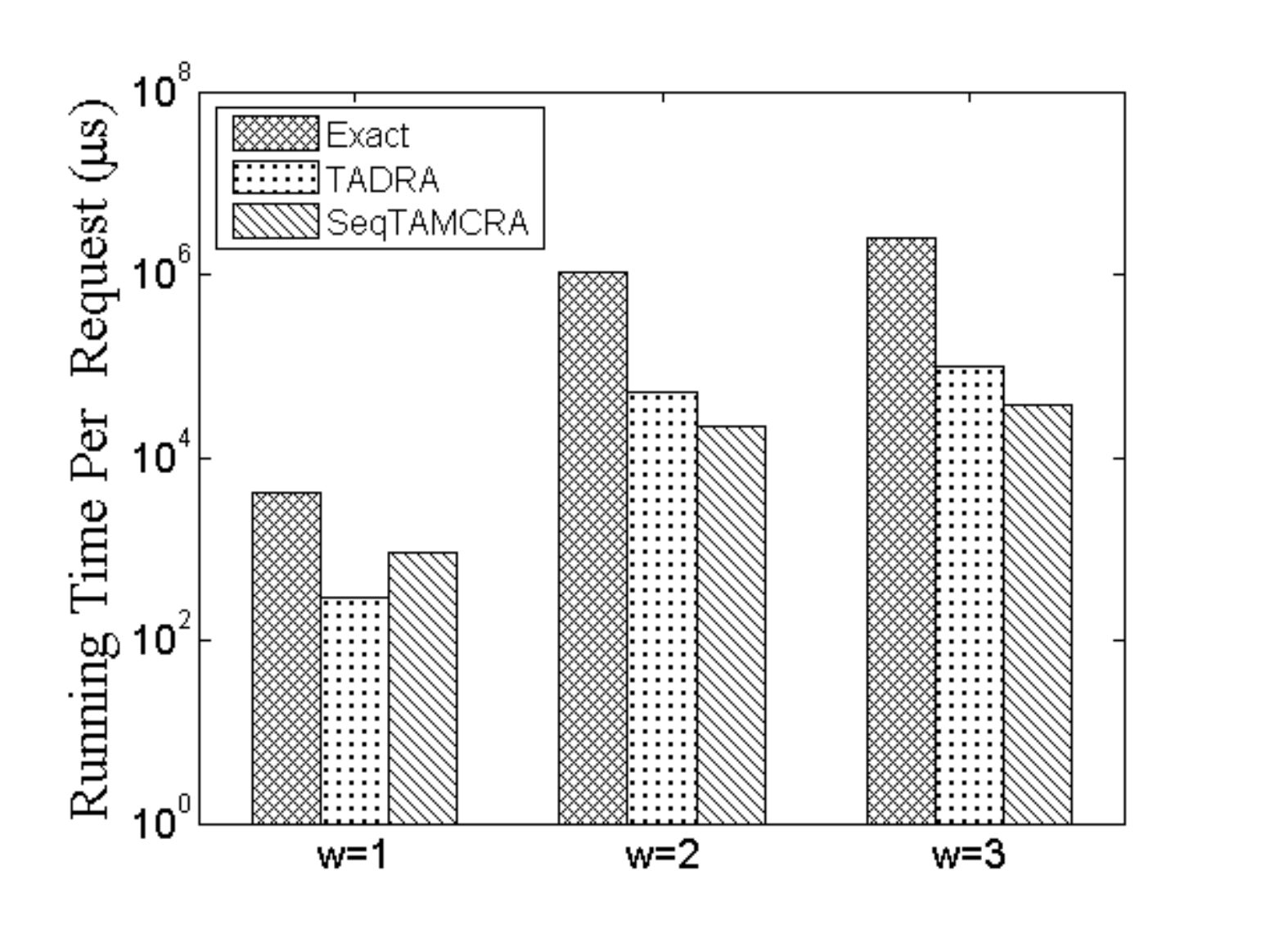}\label{Fig:TimeGEANT}}
\caption{Running time per request in two networks: (a) USANet (b) G\'{E}ANT .}%
\label{Fig:ABDCR_Time}%
\end{figure}

When the solution does not exist, the exact algorithm needs much longer time to finish, we therefore only show the algorithms' running times (in log scale) for all their accepted requests in Fig.~\ref{Fig:ABDCR_Time}. Even in this case, we see that the exact algorithm is still more time consuming than the others. TADRA requires more running time than SeqTAMCRA when $w>1$, since its graph (input) size increases $w$ times. 

\section{Conclusion} \label{Sec:Conclusion}

In this paper, we have first studied the Reliable VM Placement (RVMP) problem. We have shown that the RVMP problem is NP-hard, and cannot be approximated to an arbitrary degree. To solve it, we have proposed an exact INLP as well as an efficient heuristic, and compare these algorithms with another two heuristic modifications. The simulation results reveal that, our proposed heuristic can always achieve a better performance in terms of acceptance ratio and the number of used nodes than the modified heuristics, although it consumes an (acceptably) higher running time. On the other hand, the exact INLP can always achieve the best performance, but its running time is significantly larger than all the heuristics. 
Following that, we have studied the Availability-Based Delay-Constrained Routing (ABDCR) problem. We have shown that the ABDCR problem is NP-hard and have proposed both an exact algorithm and two heuristics to solve it. Finally, we have tested these 3 algorithms via simulations on two networks. The simulation results indicate that the exact algorithm can always achieve the highest acceptance ratio, but this comes at the expense of much higher running time. Meanwhile, SeqTAMCRA and TADRA return close to optimal result in a shorter time for $w=1$ and $w>1$, respectively, which suggests a combination of use when the computation time is a big concern. 

In reality, the cloud provider can first solve the ABDCR problem via the proposed algorithms in Section \ref{Sec:RoutingExactAlg} and/or \ref{Sec:RoutingHeu} for different node pairs in the network. Those returned solutions serve as the input for the RVMP problem. At last, the cloud provider can solve the RVMP problem by using the proposed solutions in Section \ref{Subsec:ExactAlg} and/or \ref{Subsec:Heu}. This is one possible scenario how a practitioner can face both problems in Sections \ref{Sec:ProbDef} and \ref{Sec:Routing}, and can apply our proposed solutions in the sequential order. 
In case the network (problem) size is too large but the computation time needs to be short, one possible approach is to first exclude some ``poor availability'' nodes from the graph and then run the algorithm(s) on the remainder of the network. 


%

\ifCLASSOPTIONcompsoc
  \section*{Acknowledgments}
\else
  \section*{Acknowledgment}
\fi

This research was funded by the joint EU FP7 Marie Curie Actions CleanSky Project, Contract No. 607584. 

\ifCLASSOPTIONcaptionsoff
  \newpage
\fi



%



\appendices

\section{RVMP problem under the single-node failure}
In this appendix, we provide an INLP for the RVMP problem under the single-node failure scenario. The notations and variables follow the same with the ones in Section~\ref{Subsec:ExactAlg}. Slightly different from Eqs.~(\ref{Eq:Obj})-(\ref{Eq: AvbConstraint}), Eq.~(\ref{Eq:s1PlaceConstraint}) ensures that the primary placement plan ($P^1_{vn}$) to put all the VMs on the network. 
Eq.~(\ref{Eq:s2PlaceConstraint}) ensures that the primary placement plan does not place the same VM on the same node whose availability is less than $\delta$ with the backup placement plan ($P^2_{vn}$).
Eq.~(\ref{Eq:s2AvbConstraint}) accounts for the placement availability. By setting $\beta=\frac{1}{\min_{j \in \mathcal{N}}(A_j)}$, as long as one VM $v \in V $ is placed on two nodes, the regarded ``availability'' value is returned as greater than $1$, which is also greater than $\delta$.

\textbf{Objective:}

\begin{equation}
\min \sum_{n \in \mathcal{N}} \max_{v \in V} \left(  P^1_{vn},P^2_{vn}   \right)
\end{equation}

\textbf{Placement constraint:}
\begin{flalign}
\sum_{n \in \mathcal{N}} P^1_{vn} \geq 1 ~~~\forall v \in V
\label{Eq:s1PlaceConstraint}
\end{flalign}

\begin{flalign}
P^1_{vn} \neq P^2_{vn} ~~~ \forall v \in V, n \in \mathcal{N}: A_n<\delta
\label{Eq:s2PlaceConstraint}
\end{flalign}

\textbf{Storage constraint:}

\begin{flalign}
\sum_{v \in V} \left(  \max(P^1_{vn},P^2_{vn}) \cdot c_v \right)  \leq s_n ~~~ \forall n \in \mathcal{N}
\label{Eq:s3PlaceConstraint}
\end{flalign}

\textbf{Delay and connection availability constraint:}

\begin{flalign}
F(m,n,A(m,n),T(m,n)) \cdot P^{h_1}_{am} \cdot P^{h_2}_{bn}=1   \\       
 ~~\forall h_1, h_2=1,2, (m,n) \in \mathcal{L},  a, b \in V: a \neq b   \nonumber
\end{flalign}

\textbf{VM placement availability constraint:}

\begin{flalign}
\min_{n \in \mathcal{N}}  (1-P^1_{vn}+P^1_{vn}A_n)+ \max_{n \in \mathcal{N}} (P^2_{vn} \cdot A_n) \cdot \beta \geq \delta ~  \nonumber \\
\forall v \in V,~ \beta=\frac{1}{\min_{n \in \mathcal{N}}(A_n)}
\label{Eq:s2AvbConstraint}
\end{flalign}

\section{ABDCR problem under the single-link failure }

In this section, it is assumed that at any particular time point, at most one link may fail. For a certain request $r(s, t, \eta, D)$, when $w=1$ in the ABDCR problem, we provide a polynomial-time algorithm to solve it. First, we remove the links whose availability is no greater than $\eta$. After that, we run shortest path algorithm from $s$ to $t$ to obtain a minimum delay path. If there is a feasible path and its delay is less than $D$, then it is the optimal solution, otherwise there is no solution.

When $k \geq 2$, if the optimal solution consists of $k$ fully link-disjoint paths, then $2$ fully link-disjoint paths also exist and have availability $1$ under the single-link failure scenario, which is optimal. When the optimal solution consists of $k$ partially link-disjoint paths, then $w=2$ (partially) link-disjoint paths are also enough. The reason is that the availability of partially link-disjoint paths is decided by one unprotected link (say $l$). Hence, it suffices to find $w=2$ link-disjoint paths. The proof for $w>2$ follows analogously from the proof for $w=2$. Consequently, we have the following theorem:

\begin{theorem}
The ABDCR problem for $w=2$ under the single-node failure scenario is NP-hard.
\end{theorem}

\begin{IEEEproof}
\textbf{Fully link-disjoint paths:} We regard that the link delay as the link weight in the graph. Hence, any two fully link-disjoint paths have connection availability $1$, which can always satisfy $\eta$. Now, the ABDCR problem is equivalent to the decision version of the disjoint min-max problem, which is to find two link-disjoint paths from a source to a destination, such that the maximum path weight is minimized. Since the disjoint min-max problem is NP-hard \cite{Li90}, our proof is therefore complete.

\begin{figure}[tbh]
\centering
\includegraphics[trim = 0mm 0mm 0mm 0mm,clip,width=0.35\textwidth]{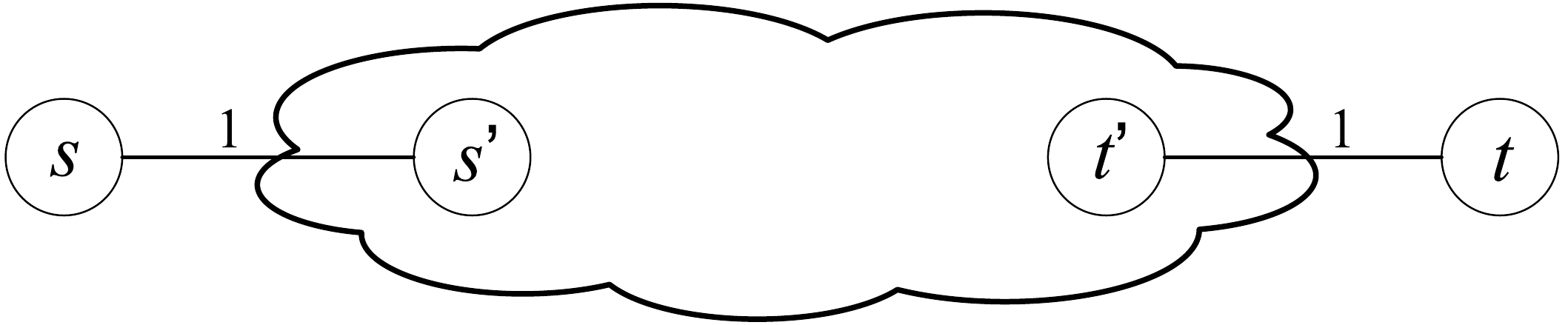}
\caption{Reduction of ABPS problem from partially link disjoint to fully
link disjoint.}%
\label{Fig: ABPS_NPhard}%
\end{figure}

\textbf{Partially link-disjoint paths:} The case for partially link-disjoint paths can be reduced to the case of fully link-disjoint paths by a transformation such as in Fig. \ref{Fig: ABPS_NPhard}. More specifically, if we assume any link in Fig. \ref{Fig: ABPS_NPhard}, except for $(s,s')$ and $(t',t)$, has availability less than $\delta$, then no link, except for $(s,s')$ and $(t',t)$, can be the unprotected link in the solution of the ABDCR problem for the partially link disjoint case from $s$ to $t$. In this context, for such $\eta$, solving the fully link-disjoint ABDCR problem from $s'$ to $t'$ is equivalent to solving the partially link-disjoint ABDCR problem from $s$ to $t$.

\end{IEEEproof}

\bibliographystyle{IEEEtran}

%

\begin{IEEEbiography}{Song Yang}
received the B.S. degree in software engineering and the M.S. degree in computer science from the Dalian University of Technology, Dalian, Liaoning, China, in 2008 and 2010, respectively, and the Ph.D. degree from Delft University of Technology, The Netherlands, in 2015. He is currently a postdoc researcher in GWDG. His research interests focus on network optimization algorithms in optical networks, stochastic networks and data center networks.
\end{IEEEbiography}
\begin{IEEEbiography}{Philipp Wieder}
is deputy leader of data center of the GWDG at the University of G{\"o}ttingen, Germany. He received his doctoral degree from TU Dortmund in Germany.
He is active in the research areas on clouds, grid and service oriented infrastructures for several years. His research interest lies in distributed system, service level agreements and resource scheduling. He has been actively involved in the FP7 IP PaaSage, SLA@SOI and SLA4D-Grid projects. 
\end{IEEEbiography}
\begin{IEEEbiography}{Ramin Yahyapour}
is full professor at the Georg-August University of G{\"o}ttingen. He is also managing director of
the GWDG, a joint compute and IT competence center of the university and the Max Planck Society. Dr. Yahyapour holds a doctoral degree in Electrical Engineering and his research interest lies in the area of efficient resource management in its application to service-oriented infrastructures, clouds, and data management. He is especially interested
in data and computing services for eScience. He gives lectures on parallel processing systems, service computing, distributed systems, cloud computing, and grid technologies. He was and is active in several national and international research projects. Ramin Yahyapour serves regularly as reviewer for funding agencies and
consultant for IT organizations. He is organizer and program committee member of conferences and workshops as well as reviewer for journals. 
\end{IEEEbiography}

\begin{IEEEbiography}
{Stojan Trajanovski} is a visiting researcher at Delft University of Technology in The Netherlands. He was a postdoctoral researcher at the University of Amsterdam. He received his PhD degree (\emph{cum laude}, 2014) from Delft University of Technology and his master degree in Advanced Computer Science (\emph{with distinction}, 2011) from the University of Cambridge, United Kingdom. He also holds an MSc degree in Software Engineering (2010) and a Dipl. Engineering degree (\emph{summa cum laude}, 2008) from Ss. Cyril and Methodius University in Skopje. He successfully participated at international science olympiads, winning a bronze medal at the International Mathematical Olympiad (IMO) in 2003.

His main research interests include network science, network robustness, complex networks, game theory, and optimization algorithms.
\end{IEEEbiography}

\begin{IEEEbiography}{Xiaoming Fu}
received his Ph.D. in computer science from Tsinghua University, Beijing, China in 2000. He was then a research staff at the Technical University Berlin until joining the University of G\"ottingen, Germany in 2002, where he has been a professor in computer science and heading the Computer Networks Group since 2007. He has spent research visits at universities of Cambridge, Uppsala, UPMC, Columbia,  UCLA, Tsinghua, Nanjing, Fudan, and PolyU of Hong Kong. 
Prof. Fu's research interests include network architectures, protocols, and applications. He is currently an editorial board member of IEEE Communications Magazine, IEEE Transactions on Network and Service Management, and Elsevier Computer Communications, and has served on the organization or program committees of leading conferences such as INFOCOM, ICNP, ICDCS, MOBICOM, MOBIHOC, CoNEXT, ICN and COSN. He is an IEEE Senior Member and an IEEE Communications Society Distinguished Lecturer. 
\end{IEEEbiography}

%




\end{document}